\documentclass[12pt]{article}
\usepackage{mathrsfs}
\usepackage{graphicx}
\usepackage{amssymb,amsmath,amsthm}
\usepackage{epstopdf}
\usepackage{bm}
\usepackage[pdftex,colorlinks,urlcolor=blue]{hyperref}
 \textwidth = 6.5 in \textheight = 9 in \oddsidemargin =
0.0 in \evensidemargin = 0.0 in \topmargin = 0.0 in \headheight =
0.0 in \headsep = 0.0 in
\def\ba{\begin{eqnarray}}
\def\ea{\end{eqnarray}}
\def\nn{\nonumber}
\def\L{\mathcal{L}}

\def\D{\mathcal{D}}

\def\:{\boldsymbol{:}}

\def\x{\mathbf{x}}

\def\p{\mathbf{p}}

\def\n{\hat{\mathbf{n}}}

\def\slash{\makebox[-.075cm][l]{/}}
\long\def\symbolfootnote[#1]#2{\begingroup%
\def\thefootnote{\fnsymbol{footnote}}\footnote[#1]{#2}\endgroup}

\begin{document}

\begin{flushright}
%Sample Draft: 
%\today
\end{flushright}
\begin{center} 
\vglue .06in
{\Large \bf {Interaction of Neutrinos with a Cosmological K-essence Scalar.\symbolfootnote[1]{This work was supported by the US department of energy.}}}\\[.5in]

{\bf Christopher S. Gauthier\symbolfootnote[2]{Electronic address: csg@umich.edu}, Ryo Saotome\symbolfootnote[5]{Electronic address: rsaotome@umich.edu}, and Ratindranath Akhoury\symbolfootnote[3]{Electronic address: akhoury@umich.edu}}\\
[.1in]
{\it{Michigan Center for Theoretical Physics\\
Randall Laboratory of Physics\\
University of Michigan\\
Ann Arbor, Michigan 48109-1120, USA}}\\[.2in]
\end{center}

%+++++++++++++++++++++++++++++++++++++++++++++++++++++++++++++++++++++++++++++++
\begin{abstract}
In this paper we study a novel means of coupling neutrinos to a Lorentz violating background k-essence field. We first look into the effect that k-essence has on the neutrino dispersion relation and derive a general formula for the neutrino velocity in the presence on a k-essence background. The influence of k-essence coupling on neutrino oscillations is then considered. It is found that a non-diagonal k-essence coupling leads to an oscillation length that goes like $\lambda \sim E^{-1}$ where $E$ is the energy. This should be contrasted with the $\lambda \sim E$ dependence seen in the standard mass-induced mechanism of neutrino oscillations. While such a scenario is not favored experimentally, it places constraints on the interactions of the neutrino with a cosmological k-essence scalar background by requiring it to be flavor diagonal. All non-trivial physical effects discussed here require the speed of sound to be different from the speed of light and hence are primarily a consequence of Lorentz violation.
\end{abstract}
%+++++++++++++++++++++++++++++++++++++++++++++++++++++++++++++++++++++++++++++++

\newpage

%+++++++++++++++++++++++++++++++++++++++++++++++++++++++++++++++++++++++++++++++
%+++++++++++++++++++++++++++++++++++++++++++++++++++++++++++++++++++++++++++++++
%+++++++++++++++++++++++++++++++++++++++++++++++++++++++++++++++++++++++++++++++
%+++++++++++++++++++++++++++++++++++++++++++++++++++++++++++++++++++++++++++++++
\section{Introduction}
%+++++++++++++++++++++++++++++++++++++++++++++++++++++++++++++++++++++++++++++++
%+++++++++++++++++++++++++++++++++++++++++++++++++++++++++++++++++++++++++++++++
%+++++++++++++++++++++++++++++++++++++++++++++++++++++++++++++++++++++++++++++++
%+++++++++++++++++++++++++++++++++++++++++++++++++++++++++++++++++++++++++++++++
The observation of the accelerated expansion of the universe \cite{Riess:1998cb,Perlmutter:1998np} has been one of the most important recent discoveries in cosmology.  Many possible explanations have been put forward which may be classified under two general classes: models with a cosmological constant, or dynamical models of dark energy. Their common feature is to provide fluids with negative pressure to drive the acceleration. Among the dynamical dark energy models, only k-essence \cite{ArmendarizPicon:2000ah}, has the 
advantage of explaining not only the current phase of accelerated expansion but also the coincidence problem, i.e., why the cross-over from the matter dominated era to the current  era happened so recently in the past. This explanation, however, is not without its own problems as was first pointed out in \cite{Bonvin:2006vc}. It was shown there that in order to solve the coincidence problem, the universe had to go through an era where the sound speed or the speed of propagation of the k-essence fluctuations must become super-luminal. This problem was addressed in \cite{Babichev:2007dw}, where it was shown that at the classical level, super-luminal propagation does not necessarily imply causal paradoxes. In particular,  propagation in a k-essence background does not have any additional causal difficulties  over general relativity where the only problems are associated with space-times which admit closed time-like curves.
In the course of this analysis, the authors found a very interesting way of describing the propagation of k-essence fluctuations in terms of an emergent metric which depends not only on the space-time metric but also the background cosmological k-essence scalar. One may thus think of this nontrivial, Lorentz violating cosmological k-essence background as the aether in which matter perturbations propagate. This emergent metric description is used in this paper to couple the k-essence background to neutrinos. 
%+++++++++++++++++++++++++++++++++++++++++++++++++++++++++++++++++++++++++++++++

%+++++++++++++++++++++++++++++++++++++++++++++++++++++++++++++++++++++++++++++++
If there is a scalar field pervading the universe then the effective field theory viewpoint implies that it must undergo interactions with the matter that is present. The question of the observabilty of  dark energy directly through its couplings to ordinary matter  is an important one \cite{Casas:1991ky,Holden:1999hm,Carroll:1998zi,Damour:1990tw,Gubser:2004du,Farrar:2003uw,Barshay:2005kd,Bertolami:2008rz,LeDelliou:2007am,Carroll:2008ub,Bean:2008ac,Bean:2007ny} and this paper attempts to address aspects of it in the context of k-essence. On short distance scales the universe is inhomogenous with, in particular, plenty of black holes. Thus the interactions with dark energy with black holes could be one way to study the above mentioned question. In this paper, however, we will be concerned only with the effect of the k-essence background aether on the propagation of neutrinos. The coupling of fermions to this background is in itself an interesting question from the theoretical point of view. Most studies of the interaction of dark energy with fermions, couple them through a Yukawa like interaction \cite{Bean:2008ac}, which is quite reasonable. However, in this paper, we do this differently using the vierbeins constructed out of the emergent metric. Throughout this article the k-essence field is treated strictly as a background, however we see no reason not to treat it as a dynamical field. As we argue in section \S\ref{BackGround}, within the effective field theory methodology, the terms with higher derivatives of the fermion fields do not give rise to ghosts. The main focus of this paper is on looking for observable consequences of dark energy, so in this case the fermion in question is the neutrino which we show undergoes flavor oscillations when traveling through the k-essence aether. However, the way we introduce the fermion-k-essence coupling could be used to obtain new types of interactions between dark energy and other forms of matter including dark matter. The emergent metric from \cite{Babichev:2007dw} which is used throughout this paper to couple the dark energy to neutrinos is covariantly constant. In a future publication we will discuss how to consider even more generalized couplings by introducing torsion in the emergent space-time.
%+++++++++++++++++++++++++++++++++++++++++++++++++++++++++++++++++++++++++++++++

%+++++++++++++++++++++++++++++++++++++++++++++++++++++++++++++++++++++++++++++++
It has been noticed previously \cite{Babichev:2007dw}, in the context of the propagation of k-essence fluctuations in a classical background that defines the emergent space-time, that Lorentz invariance is lost when the speed of propagation of the fluctuations (the speed of sound $c_s$), is different from the speed of light. The same is true for the propagation of neutrinos in this background. In fact, all of the physically interesting results that we obtain in this paper are present only when $c_s \neq c$, i.e., when there is Lorentz violation. Nontrivial neutrino flavor oscillations require, in addition, non-diagonal flavor couplings of neutrinos to the k-essence background. In the past various models of neutrino oscillations  have been considered which require an explicit violation of Lorentz invariance \cite{Coleman:1997xq} or of the equivalence principle, \cite{Gasperini:1988zf,Halprin:1991gs}. The energy dependence of the oscillation length is the same in these models as the one considered in this paper. In this sense our model may be considered a theoretically and phenomenologically motivated manifestation of the same phenomenon. We should emphasize that we do not have any violation of the equivalence principle, but the emergent metric which contains contributions from the k-essence background can be different for different flavors of neutrinos. This is made possible in our model by a flavor non-diagonal coupling in the part of the emergent metric involving the k-essence background only.
%+++++++++++++++++++++++++++++++++++++++++++++++++++++++++++++++++++++++++++++++

%+++++++++++++++++++++++++++++++++++++++++++++++++++++++++++++++++++++++++++++++
K-essence is a theory with noncanonical kinetic terms and coupling it to neutrinos through the vierbein of the emergent metric alters the speed of propagation of the neutrinos. The consequent dispersion relations are analyzed in section \S\ref{BackGround}. The data from supernova 1987 is then used as an input to constrain some of the parameters of our model. In section \S\ref{NeutrinoOscillations} we consider the possibility of neutrino oscillations induced by their coupling to the k-essence background. The more interesting and novel case is when the different neutrino flavors couple with different strengths to the k-essence field. Oscillations are induced essentially due to the fact that the speeds of propagation of the different neutrino species in the background aether are consequently different. In this case neutrinos would oscillate even if they were massless. In section \S\ref{NonDiagCase} we discuss this case in some detail and obtain a general formula for the oscillation probability with massive neutrinos. Our results are quantitatively different than the case of flavor oscillations with only massive neutrinos. In particular, the oscillation length varies with the inverse power of the neutrino energy. Such a behavior is ruled out by the data from the Kamiokande experiment \cite{Fukuda:1998mi}. Thus, we are able to place bounds on the allowed strengths of the k-essence coupling to neutrinos. In particular, the data strongly favor diagonal flavor couplings of k-essence to neutrinos. As a preliminary to this analysis, in section \S\ref{DiagCase} we discuss the case of neutrino oscillations with massive neutrinos but with equal coupling strengths of all flavors to the k-essence background. Here we find a rather simple modification of the well known formula for the flavor oscillations of massive neutrinos with the only difference arising due to the fact that the neutrinos travel along geodesics in the emergent space-time. In section \S\ref{Conclusions} we present our conclusions. Certain technical details of the coupling of Dirac fermions to the emergent metric are relegated to an appendix.
%+++++++++++++++++++++++++++++++++++++++++++++++++++++++++++++++++++++++++++++++

%+++++++++++++++++++++++++++++++++++++++++++++++++++++++++++++++++++++++++++++++
%+++++++++++++++++++++++++++++++++++++++++++++++++++++++++++++++++++++++++++++++
%+++++++++++++++++++++++++++++++++++++++++++++++++++++++++++++++++++++++++++++++
%+++++++++++++++++++++++++++++++++++++++++++++++++++++++++++++++++++++++++++++++
\section{Neutrino Coupling To a K-essence Background}
\label{BackGround}
%+++++++++++++++++++++++++++++++++++++++++++++++++++++++++++++++++++++++++++++++
%+++++++++++++++++++++++++++++++++++++++++++++++++++++++++++++++++++++++++++++++
%+++++++++++++++++++++++++++++++++++++++++++++++++++++++++++++++++++++++++++++++
%+++++++++++++++++++++++++++++++++++++++++++++++++++++++++++++++++++++++++++++++
Before we can talk about neutrino interactions with k-essence we should first review the latter \cite{ArmendarizPicon:2000ah,Babichev:2007dw}. In general k-essence is a theory of a scalar field with non-canonical kinetic terms. The lagrangian of a single k-essence scalar $\phi$ is usually denoted by a single function $L(X,\phi)$ where $X = \frac{1}{2} \nabla_{\mu}\phi \nabla^{\mu}\phi$. For a given solution $\phi$ to the k-essence equations of motion, the behavior of perturbations $\chi = \delta \phi$ in the k-essence field around the background $\phi$ can be written can be described by a canonical scalar field action, but with the space-time metric $g_{\mu\nu}$ replaced by an emergent space-time metric $G_{\mu\nu}$ given by \cite{Babichev:2007dw},  
%+++++++++++++++++++++++++++++++++++++++++++++++++++++++++++++++++++++++++++++++
\begin{gather}
G_{\mu\nu}
=
\Omega^{-2}
\left(
g_{\mu\nu}
+
\frac{c_{s}^{2} - 1}{2 X}
\nabla_{\mu}
\phi
\nabla_{\nu}
\phi
\right)
\end{gather}
%+++++++++++++++++++++++++++++++++++++++++++++++++++++++++++++++++++++++++++++++
where $\Omega^{2} = \frac{c_{s}}{L_{X}}$\footnote{Derivatives of $L$ with respect to $X$ are denoted by a subscript, so that $L_{X} = \frac{\partial L}{\partial X}$ and $L_{XX} = \frac{\partial^{2} L}{\partial X^{2}}$.} and $c_{s}$ is the sound speed of k-essence fluctuations,which is defined as 
%+++++++++++++++++++++++++++++++++++++++++++++++++++++++++++++++++++++++++++++++
\begin{gather}
c_{s}^{2} = \left(\frac{\partial p_{k}}{\partial \rho_{k}}\right)_{\phi}
\label{CSSQU}
\end{gather}
%+++++++++++++++++++++++++++++++++++++++++++++++++++++++++++++++++++++++++++++++
Here $\rho_{k}$ and $p_{k}$ denote the energy density and pressure of the k-essence background, and the subscript $\phi$ signals that (\ref{CSSQU}) should be evaluated while holding $\phi$ constant. In the case where $\nabla_{\mu}\phi$ is a time-like vector (i.e. $X>0$) then it can be shown that $p_{k} = L$ and $\rho_{k} = 2 X L_{X} - L$, and therefore it follows that
%+++++++++++++++++++++++++++++++++++++++++++++++++++++++++++++++++++++++++++++++
\begin{gather}
c_{s}^{2}
=
\frac{L_{X}}{L_{X} + 2 X L_{XX}} 
\end{gather}
%+++++++++++++++++++++++++++++++++++++++++++++++++++++++++++++++++++++++++++++++
Because $g_{\mu\nu}$ is replaced by $G_{\mu\nu}$ in the $\chi$ field action, the characteristics of $\chi$ follow the geodesics of $G_{\mu\nu}$ and not $g_{\mu\nu}$. This interesting fact implies that $\chi$ has a different  causal structure then all other fields. In particular, if in some frame $g_{\mu\nu}= \eta_{\mu\nu}$ and the background k-essence field is uniform, then the emergent space-time metric is given by 
%+++++++++++++++++++++++++++++++++++++++++++++++++++++++++++++++++++++++++++++++
\begin{gather}
G_{\mu\nu}
dx^{\mu}
dx^{\nu}
\propto
c_{s}^{2}
dt^{2}
-
d\x^{2}
\label{kessenceMetric}
\end{gather}
%+++++++++++++++++++++++++++++++++++++++++++++++++++++++++++++++++++++++++++++++
The metric $G_{\mu\nu}$ defines a different causal structure than $g_{\mu\nu}$. The ``light'' cones of $G_{\mu\nu}$ are defined by characteristics with velocities $c_{s}$ instead of the speed of light. It was shown in \cite{Babichev:2007dw} that even if $c_{s}$ exceeded the speed of light, causality in a k-essence theory would still be preserved despite the superluminal speed of k-essence fluctuations. This fact can be roughly understood by thinking of the k-essence metric (\ref{kessenceMetric}) as the space-time interval in the special relativity but with a different value for the speed of light. Therefore causality is preserved in k-essence for much the same reason that it is preserved in special relativity. However, in order for this rational to hold the k-essence action most satisfy certain constraints. These constraints arise from the need to make $G_{\mu\nu}$ have the proper signature. This translates to the requirement that the k-essence action satisfies
%+++++++++++++++++++++++++++++++++++++++++++++++++++++++++++++++++++++++++++++++
\begin{gather}
1 + \frac{2 X L_{XX}}{L_{X}}
>0.
\label{SigCon}
\end{gather}
%+++++++++++++++++++++++++++++++++++++++++++++++++++++++++++++++++++++++++++++++
Note that this condition is equivalent to the stability constraint: $c_{s}^{2}>0$.
%+++++++++++++++++++++++++++++++++++++++++++++++++++++++++++++++++++++++++++++++

In this paper we will take inspiration from the k-essence perturbation action, and consider the possibility of other fields coupling to $G_{\mu\nu}$. In particular, we will take the action of a neutrino coupled to a gravitational metric $g_{\mu\nu}$ and replace this with the k-essence metric $G_{\mu\nu}$. The action of our hypothetical k-essence-coupled (Dirac) neutrino $\nu$ is given by:  
%+++++++++++++++++++++++++++++++++++++++++++++++++++++++++++++++++++++++++++++++
\begin{gather}
S
=
\int d^{4}x
E
\bar{\nu}
\left[
i
\tilde{\gamma}^{\mu}
\D_{\mu}
-
M
\right]
\nu
\label{NeutrinoAction}
\end{gather}
%+++++++++++++++++++++++++++++++++++++++++++++++++++++++++++++++++++++++++++++++
where $E = \det E_{\mu}^{a}$ and $\tilde{\gamma}^{\mu} = E^{\mu}_{a} \gamma^{a}$, and $\gamma^{a}$ are the standard gamma matrices. The vierbein field $E^{\mu}_{a}$ of the emergent space-time geometry and its inverse $E_{\mu}^{a}$ are given by
%+++++++++++++++++++++++++++++++++++++++++++++++++++++++++++++++++++++++++++++++
\begin{gather}
E_{a}^{\mu}
=
\Omega
\left(
e^{\mu}_{a}
+
\frac{g u}{2 X}
e^{\rho}_{a}
\nabla_{\rho}\phi
\nabla^{\mu}
\phi
\right)
,
\quad
E^{a}_{\mu}
=
\Omega^{-1}
\left(
e^{a}_{\mu}
+
\frac{g \bar{u}}{2 X}
e^{\rho a}
\nabla_{\rho}\phi
\nabla_{\mu}
\phi
\right).
\label{EmergentVierbein}
\end{gather}
%+++++++++++++++++++++++++++++++++++++++++++++++++++++++++++++++++++++++++++++++
Here we have defined $u = \frac{1}{c_{s}} - 1$ and $\bar{u} = - \frac{u}{1 + g u}$\footnote{Note, that in our definition of $u$ and $\bar{u}$ we have assumed that the 4-gradient of the k-essence background field is time-like. In the case of space-like $\nabla_{\mu}\phi$, the sound speed is given by the inverse of the expression for the sound speed in the time-like case. Thus, for space-like $\nabla_{\mu}\phi$ one should replace $c_{s} \rightarrow \frac{1}{c_{s}}$ in the definition of $u$ and $\bar{u}$.}. Further note that $\nabla^{\mu}\phi = g^{\mu\nu}\nabla_{\nu}\phi$.  Also, we have included a coupling constant $g$, which accounts for the interaction strength of the k-essence background to the neutrinos. The emergent metric $G_{\mu\nu}$ is still covariantly constant as in \cite{Babichev:2007dw}.

At this point we would like to emphasize two features of the action. First, we note that when $c_s \neq 1$ the model is not invariant under Lorentz transformations \cite{Babichev:2007dw}. As we will see in detail in section \S\ref{NeutrinoOscillations}, all the physical effects that we discuss in this paper are consequences of this Lorentz violation in the sense that they vanish at $c_s = 1$. Other features like nondiagonal flavor couplings are also important to get nontrivial flavor oscillations, however, Lorentz violations must always be present.
Secondly, in this model we treat the k-essence background as a classical field, which does not experience any appreciable back reaction effects from the neutrino field. This allows us to treat the neutrino/k-essence coupling as a contribution to the kinetic term in the neutrino action. If the k-essence field was dynamical, this would lead to higher order derivatives of the neutrino field in the k-essence equation of motion, which could potentially create ghosts in the quantum theory. This is not a problem for this paper since we treat the k-essence scalar strictly as a classical background, however, we would like to emphasize that within the effective field theory methodology treating the k-essence scalar as dynamical would not give rise to such problems in any case. The terms with higher derivatives of the fermion field would be considered as higher order in the low energy effective action expansion.

The derivative operator $\D_{\mu}$ in (\ref{NeutrinoAction}) represents the spinor covariant derivative with respect to the emergent k-essence background described by $E^{\mu}_{a}$. The proper definition of $\D_{\mu}$ and its specific form in the case of a general k-essence field in a flat space-time background are given in appendix \ref{SpinCon}. From here on out we will ignore the spinor connection term in $\D_{\mu}$, which we justify on the basis that higher derivatives of the k-essence field are negligibly small at the present time in most models of k-essence. However, if at one time in the history of the universe the connection was not small, its effect can be accounted for by replacing $M^{2} \rightarrow M^{2} + R/4$, where $R$ is the scalar curvature of vierbein $E_{\mu}^{a}$. The emergent space-time metric is by definition given by $G_{\mu\nu} = E_{\mu}^{a}E_{\nu}^{b}\eta_{ab}$, which is
%+++++++++++++++++++++++++++++++++++++++++++++++++++++++++++++++++++++++++++++++
\begin{gather}
G_{\mu\nu}
=
E_{\mu}^{a}E_{\nu}^{b}\eta_{ab}
=
\Omega^{-2}
\left(
g_{\mu\nu}
+
\frac{\bar{w}}{2 X}
\nabla_{\mu}\phi
\nabla_{\nu}
\phi
\right)
\label{EmergentG}
\end{gather}
%+++++++++++++++++++++++++++++++++++++++++++++++++++++++++++++++++++++++++++++++
where $\bar{w} = 2 g \bar{u} + g^{2} \bar{u}^{2}$. Note that if $g=1$ then $\bar{w} = c_{s}^{2} - 1$, which is the standard result for the emergent metric in k-essence theories. The inverse of $G_{\mu\nu}$ is given by
%+++++++++++++++++++++++++++++++++++++++++++++++++++++++++++++++++++++++++++++++
\begin{gather}
G^{\mu\nu}
=
\Omega^{2}
\left(g^{\mu\nu}  
+
\frac{w}{2 X}
\nabla^{\mu}\phi
\nabla^{\nu}\phi
\right)
\end{gather}
%+++++++++++++++++++++++++++++++++++++++++++++++++++++++++++++++++++++++++++++++
where $w = 2 g u + g^{2} u^{2} = - \frac{\bar{w}}{1 + \bar{w}}$. The determinant of $G_{\mu\nu}$ is given by 
%+++++++++++++++++++++++++++++++++++++++++++++++++++++++++++++++++++++++++++++++
\begin{gather}
E^{2}
=
-\det G_{\mu\nu}
=
-(\det g_{\mu\nu})
\frac{\Omega^{-8}}{1 + w}
\end{gather}
%+++++++++++++++++++++++++++++++++++++++++++++++++++++++++++++++++++++++++++++++
In order for this modified k-essence induced metric to have the proper signature: $1 + w >  0$. Note that this condition is equivalent to (\ref{SigCon}) in the case when $g=1$. The k-essence coupled Dirac equation reads
%+++++++++++++++++++++++++++++++++++++++++++++++++++++++++++++++++++++++++++++++
\begin{gather}
\left(
i \tilde{\gamma}^{\mu}
\D_{\mu}
-
M
\right)
\nu(t,x)
=0
\end{gather}
%+++++++++++++++++++++++++++++++++++++++++++++++++++++++++++++++++++++++++++++++
If we square this equation we can obtain a Klein-gordon equation for $\nu(t,x)$
%+++++++++++++++++++++++++++++++++++++++++++++++++++++++++++++++++++++++++++++++
\begin{gather}
(
G^{\mu\nu}
\nabla_{\mu}
\nabla_{\nu}
+
R/4
+
M^{2})
\nu(t,x)
=0
\label{KleinGordon}
\end{gather}
%+++++++++++++++++++++++++++++++++++++++++++++++++++++++++++++++++++++++++++++++
Here, $R$ is the curvature scalar whose definition in terms of the connection form is 
%+++++++++++++++++++++++++++++++++++++++++++++++++++++++++++++++++++++++++++++++
\begin{gather}
R
=
E^{\mu}_{a}
E^{\nu}_{b}
\left[
\partial_{\mu}
\Omega^{ab}_{\,\,\,\,\,\, \nu}
-
\partial_{\nu}
\Omega^{ab}_{\,\,\,\,\,\, \mu}
+
\Omega^{a}_{c\mu}
\Omega^{cb}_{\,\,\,\,\,\, \nu}
-
\Omega^{a}_{c\nu}
\Omega^{cb}_{\,\,\,\,\,\, \mu}
\right]
\end{gather}
%+++++++++++++++++++++++++++++++++++++++++++++++++++++++++++++++++++++++++++++++
It is evident from (\ref{KleinGordon}) that the curvature scalar acts as a mass term. However, as we discussed earlier, in most k-essence models higher derivatives of the field $\phi$ will be negligible, and therefore we can ignore $R$ from here on out. By taking the plane wave approximation for the neutrino field the phase of $\nu$ is proportional to $e^{- i \int  p_{\mu} dx^{\nu}}$. If we assume that the interaction of the neutrino field is weak and the background geometry is flat, then the dominant space-time dependence of the neutrino field comes from the phase factor. Thus the Klein-Gordon equation in momentum-space leads to the dispersion relation
%+++++++++++++++++++++++++++++++++++++++++++++++++++++++++++++++++++++++++++++++
\begin{gather}
G^{\mu\nu}
p_{\mu}
p_{\nu}
-
M^{2}
=0
\end{gather}
%+++++++++++++++++++++++++++++++++++++++++++++++++++++++++++++++++++++++++++++++
Define an effective momentum $\tilde{p}_{\mu}$ as
%+++++++++++++++++++++++++++++++++++++++++++++++++++++++++++++++++++++++++++++++
\begin{gather}
\tilde{p}_{\mu}
=
\Omega^{-1}
e^{a}_{\mu}
E^{\nu}_{a}
p_{\nu}
=
p_{\mu}
+
\frac{g u}{2 X}
(p_{\nu}
\nabla^{\nu}\phi)
\nabla_{\mu}\phi
\end{gather}
%+++++++++++++++++++++++++++++++++++++++++++++++++++++++++++++++++++++++++++++++
The covariant and contravariant effective momentum are defined with respect to the space-time metric and not the emergent k-essence metric  $G_{\mu\nu}$. Thus the index on $\tilde{p}_{\mu}$ is raised and lowered using the space-time metric $g_{\mu\nu}$. Because of this property, it follows that
%+++++++++++++++++++++++++++++++++++++++++++++++++++++++++++++++++++++++++++++++
\begin{gather}
\Omega^{2}
\tilde{p}_{\mu}
\tilde{p}^{\mu}
=
\Omega^{2}
g^{\mu\nu}
\tilde{p}_{\mu}
\tilde{p}_{\nu}
=
g^{\mu\nu}
(e^{a}_{\mu}
E^{\rho}_{a}
p_{\rho})
(e^{b}_{\nu}
E^{\lambda}_{b}
p_{\lambda})
=
\eta^{a b}
E^{\mu}_{a}
E^{\nu}_{b}
p_{\mu}
p_{\nu}
=
G^{\mu\nu}
p_{\mu}
p_{\nu}
\end{gather}
%+++++++++++++++++++++++++++++++++++++++++++++++++++++++++++++++++++++++++++++++
For the purposes of this paper we will assume that the background space time is flat so that $g_{\mu\nu} = \eta_{\mu\nu}$ and $e^{\mu}_{a} = \delta^{\mu}_{a}$. This is in fact, a good approximation cosmologically for the applications we have in mind in this paper.Therefore, the dispersion relation in terms of the effective momentum is
%+++++++++++++++++++++++++++++++++++++++++++++++++++++++++++++++++++++++++++++++
\begin{gather}
\tilde{p}_{0}
\tilde{p}^{0}
+
\tilde{p}_{i}
\tilde{p}^{i}
=m^{2}
\label{MassShellwMass}
\end{gather}
%+++++++++++++++++++++++++++++++++++++++++++++++++++++++++++++++++++++++++++++++
where $m^{2} = \Omega^{-2} M^{2} $ is an effective mass that we have defined here for convenience. We wish to use the on mass shell condition above to find the particle velocity $v$ of the neutrinos which is represented by the group velocity $v = \frac{\partial p_{0}}{\partial |\p|}$. In general we will find that the velocity of a neutrino depends on the energy which stands in contrast to the case of a free massless neutrino, where the velocity is energy independent.

Throughout this paper we will define $p_{\mu} = (E , - \p)$, $p = |\p|$, $\p= p \n$, and $\dot{\phi} = \nabla_{0} \phi$. In the next few sections we will find the neutrino velocity for two special cases: a uniform k-essence field, and a static k-essence field, after which we will derive the velocity assuming the most general k-essence field configuration.

%+++++++++++++++++++++++++++++++++++++++++++++++++++++++++++++++++++++++++++++++
%+++++++++++++++++++++++++++++++++++++++++++++++++++++++++++++++++++++++++++++++
%+++++++++++++++++++++++++++++++++++++++++++++++++++++++++++++++++++++++++++++++
%+++++++++++++++++++++++++++++++++++++++++++++++++++++++++++++++++++++++++++++++
\subsection{Simple Case: $\phi$ is Uniform} 
%+++++++++++++++++++++++++++++++++++++++++++++++++++++++++++++++++++++++++++++++
%+++++++++++++++++++++++++++++++++++++++++++++++++++++++++++++++++++++++++++++++
%+++++++++++++++++++++++++++++++++++++++++++++++++++++++++++++++++++++++++++++++
%+++++++++++++++++++++++++++++++++++++++++++++++++++++++++++++++++++++++++++++++
Before we try and find the expression for the neutrino velocity with the most general k-essence field, let's find the velocity when $\phi$ is uniform. This is probably the most relevant case since most k-essence theories, in particular those that attempt to address the cosmological constant problem, assume that the spatial derivatives of the k-essence field are negligible compared to its time derivative  \cite{Bonvin:2006vc,ArmendarizPicon:2000ah,Malquarti:2003nn,Scherrer:2004au}. If $\phi$ is uniform then $\nabla_{i} \phi =0$, and thus
%+++++++++++++++++++++++++++++++++++++++++++++++++++++++++++++++++++++++++++++++
\begin{gather}
\tilde{p}_{0}
\tilde{p}^{0}
+
\tilde{p}_{i}
\tilde{p}^{i}
=
m^{2}
\quad
\Rightarrow
\quad
E
=
\frac{\sqrt{p^{2} + m^{2}}}{1
+
\frac{g u \dot{\phi}^{2}}{2 X}}
\end{gather}
%+++++++++++++++++++++++++++++++++++++++++++++++++++++++++++++++++++++++++++++++
By definition $X = \frac{1}{2} \dot{\phi}^{2}$. Therefore
%+++++++++++++++++++++++++++++++++++++++++++++++++++++++++++++++++++++++++++++++
\begin{gather}
v_{p}
=
\frac{E}{p}
=
c_{\nu} \sqrt{1 + \frac{m^{2}}{p^{2}}}
\label{PhaseVelocityUniform}
\end{gather}
%+++++++++++++++++++++++++++++++++++++++++++++++++++++++++++++++++++++++++++++++
and the group velocity, which represents the neutrino particle velocity is
%+++++++++++++++++++++++++++++++++++++++++++++++++++++++++++++++++++++++++++++++
\begin{gather}
v_{g}
=
\frac{\partial E}{\partial p}
=
\frac{c_{\nu}}{ \sqrt{1 + \frac{m^{2}}{p^{2}} }}
\label{GroupVelocityUniform}
\end{gather}
%+++++++++++++++++++++++++++++++++++++++++++++++++++++++++++++++++++++++++++++++
where $c_{\nu} = \frac{1}{1+ g u} = \frac{c_{s}}{(1-g)c_{s} + g}$. The speed $c_{\nu}$ plays the same role for massless neutrinos that the sound speed $c_{s}$ does for massless k-essence perturbations. Both $c_{\nu}$, and $c_{s}$ represent limiting speeds that neutrinos and k-essence perturbations, respectively, are required to not exceed as measured from the frame in which the k-essence background is uniform. The fact that in general $c_{\nu} \neq c_{s}$ is due to our inclusion of an arbitrary coupling parameter $g$. Phenomenologically this coupling is small, so we easily see that if $c_s > 1$, then so is $c_{\nu}$.

If $m^{2} =0$ then the neutrino velocity (\ref{GroupVelocityUniform}) will be equal to $c_{\nu}$.  As a massless particle coupled to the emergent background geometry, the neutrino will travel on the null geodesics of $G_{\mu\nu}$ not $g_{\mu\nu}$. In the uniform case, the emergent metric (\ref{EmergentG}) with arbitrary $g$ is
%+++++++++++++++++++++++++++++++++++++++++++++++++++++++++++++++++++++++++++++++
\begin{gather}
G_{\mu\nu}
dx^{\mu}
dx^{\nu}
\propto
c_{\nu}^{2}
dt^{2}
-
d\x^{2}
\end{gather}
%+++++++++++++++++++++++++++++++++++++++++++++++++++++++++++++++++++++++++++++++
As one can see here, null lines in $G_{\mu\nu}$ travel at a speed $c_{\nu}$. In this sense the emergent geometry of a uniform k-essence field acts just as a minkowski space-time except that limiting speed is now $c_{\nu}$ instead of the speed of light. 
%+++++++++++++++++++++++++++++++++++++++++++++++++++++++++++++++++++++++++++++++

%+++++++++++++++++++++++++++++++++++++++++++++++++++++++++++++++++++++++++++++++
%+++++++++++++++++++++++++++++++++++++++++++++++++++++++++++++++++++++++++++++++
%+++++++++++++++++++++++++++++++++++++++++++++++++++++++++++++++++++++++++++++++
%+++++++++++++++++++++++++++++++++++++++++++++++++++++++++++++++++++++++++++++++
\subsection{Slightly Less Simple Case: $\phi$ is Static} 
%+++++++++++++++++++++++++++++++++++++++++++++++++++++++++++++++++++++++++++++++
%+++++++++++++++++++++++++++++++++++++++++++++++++++++++++++++++++++++++++++++++
%+++++++++++++++++++++++++++++++++++++++++++++++++++++++++++++++++++++++++++++++
%+++++++++++++++++++++++++++++++++++++++++++++++++++++++++++++++++++++++++++++++
In direct contrast to the last case let's consider what happens to the neutrino velocity when $\phi$ is time independent, but has non-zero spatial gradients. In this case $\nabla_{\mu}\phi$ is a space-like vector (i.e. $X<0$) and the energy density and pressure are instead given by $\rho = - L$ and $p = L - 2 X L_{X}$. It follows that $c_{s}^{2} = \frac{L_{X} + 2 X L_{XX}}{L_{X}}$, which means that the definition of the sound speed for a space-like k-essence field is the inverse of sound speed for a time-like k-essence field. Thus we make the replacement $c_{s} \rightarrow \frac{1}{c_{s}}$. It follows therefore that $u = c_{s} - 1$ now. We should note, however, that the metric has not changed, only the definition of $c_{s}$. If we assume a static but spatially varying k-essence field, then the on mass shell condition states that
%+++++++++++++++++++++++++++++++++++++++++++++++++++++++++++++++++++++++++++++++
\begin{gather}
\tilde{p}_{0}
\tilde{p}^{0}
+
\tilde{p}_{i}
\tilde{p}^{i}
=m^{2}
\quad
\Rightarrow
\quad
E^{2}
=
p^{2}
\left[
1
+
w
\cos^{2}
\theta
\right]
+
m^{2}
\end{gather}
%+++++++++++++++++++++++++++++++++++++++++++++++++++++++++++++++++++++++++++++++
where $\theta$ is the angle between $\p$ and $\nabla \phi$. Note that in this case $X = - \frac{1}{2} |\nabla \phi|^{2}$. The neutrino phase velocity is then given by
%+++++++++++++++++++++++++++++++++++++++++++++++++++++++++++++++++++++++++++++++
\begin{gather}
v_{p}
=
\frac{E}{p}
=
\sqrt{1 + w \cos^{2}\theta + \frac{m^{2}}{p^{2}}}
\end{gather}
%+++++++++++++++++++++++++++++++++++++++++++++++++++++++++++++++++++++++++++++++
and the group velocity is
%+++++++++++++++++++++++++++++++++++++++++++++++++++++++++++++++++++++++++++++++
\begin{gather}
v_{g}
=
\frac{1 + w \cos^{2}\theta}{\sqrt{1 + w \cos^{2}\theta + \frac{m^{2}}{p^{2}}}}
\label{GroupVelocityStatic}
\end{gather}
%+++++++++++++++++++++++++++++++++++++++++++++++++++++++++++++++++++++++++++++++
It is informative to evaluate (\ref{GroupVelocityStatic}) at the two extremes of $\cos^{2}\theta$; that is when $\p$ and $\nabla\phi$ are parallel and when they are perpendicular. If $\p$ and $\nabla\phi$ are parallel then the angle $\theta$ between them vanishes and we find that the neutrino velocity is
%+++++++++++++++++++++++++++++++++++++++++++++++++++++++++++++++++++++++++++++++
\begin{gather}
v(\theta=0)
=
\frac{1+g u}{\sqrt{1 + \frac{1}{(1+g u)^{2} }\frac{m^{2}}{p^{2}}}}
=
\frac{c_{\nu}}{\sqrt{1 + \frac{1}{c_{\nu}^{2}}\frac{m^{2}}{p^{2}}}}
\label{GroupVelocityStaticParallel}
\end{gather}
%+++++++++++++++++++++++++++++++++++++++++++++++++++++++++++++++++++++++++++++++
where now $c_{\nu} = 1 + g u = c_{s} g- g +1$. Again, if $g =1$ then $c_{\nu} = c_{s}$. We can see that the velocity of the neutrino (\ref{GroupVelocityStaticParallel}) is almost the same as the formula given in (\ref{GroupVelocityUniform}) except that the mass term in the denominator now has a factor of $\frac{1}{c_{\nu}^{2}}$. Again if the effective mass of the neutrino is zero then the neutrino velocity is equal to the sound speed of k-essence fluctuations. This is due to the fact that the neutrino propagates on null geodesics in the emergent k-essence background.  Now, on the other hand, if $\p$ and $\nabla \phi$ are perpendicular then
%+++++++++++++++++++++++++++++++++++++++++++++++++++++++++++++++++++++++++++++++
\begin{gather}
v(\theta=\pi/2)
=
\frac{1}{\sqrt{1 + \frac{m^{2}}{p^{2}}}}
\end{gather}
%+++++++++++++++++++++++++++++++++++++++++++++++++++++++++++++++++++++++++++++++
and the formula for the velocity of the neutrino is the same as it would be if in the absence of a k-essence field. This is because in the static field case the coupling between the neutrino and k-essence is proportional to $\p\cdot\nabla\phi$. This means that the neutrino will act as a free particle propagating on a flat lorentzian space-time whenever it is traveling perpendicular to the direction of the field gradient. The directional dependence of the neutrino velocity in a spatially varying k-essence field stands in stark contrast to the neutrino velocity in uniform k-essence. If neutrinos traveling from a distant galaxy were to travel through a region of spatially varying but static k-essence field (a k-essence halo) on their way to a detector on earth, we should expect to see evidence of an-isotropy. However, even if it were possible to detect a sufficiently large neutrino flux, it is expected that any spatial variation of the k-essence field will be very small compared with its variation in time. Therefore any anisotropy in the neutrino velocity would most likely be unobservable.
%+++++++++++++++++++++++++++++++++++++++++++++++++++++++++++++++++++++++++++++++

%+++++++++++++++++++++++++++++++++++++++++++++++++++++++++++++++++++++++++++++++
%+++++++++++++++++++++++++++++++++++++++++++++++++++++++++++++++++++++++++++++++
%+++++++++++++++++++++++++++++++++++++++++++++++++++++++++++++++++++++++++++++++
\subsection{Neutrino Velocity In a General K-essence Background}
\label{NonStaticNonUniformVelocity}
%+++++++++++++++++++++++++++++++++++++++++++++++++++++++++++++++++++++++++++++++
%+++++++++++++++++++++++++++++++++++++++++++++++++++++++++++++++++++++++++++++++
%+++++++++++++++++++++++++++++++++++++++++++++++++++++++++++++++++++++++++++++++
%+++++++++++++++++++++++++++++++++++++++++++++++++++++++++++++++++++++++++++++++
Without making any assumptions about the nature of the k-essence field, the on mass shell condition (\ref{MassShellwMass}) becomes
%+++++++++++++++++++++++++++++++++++++++++++++++++++++++++++++++++++++++++++++++
\begin{gather}
(1 + \frac{w \dot{\phi}^{2} }{2 X})
\frac{E^{2}}{p^{2}}
+
\frac{w}{X}
\dot{\phi} (\n \cdot \nabla \phi)
\frac{E}{p}
+
\frac{w}{2 X}
(\n \cdot \nabla \phi)^{2}
-1
=\frac{m^{2}}{p^{2}}
\end{gather}
%+++++++++++++++++++++++++++++++++++++++++++++++++++++++++++++++++++++++++++++++
The solution for the $\frac{E}{p}$ is  
%+++++++++++++++++++++++++++++++++++++++++++++++++++++++++++++++++++++++++++++++
\begin{gather}
\frac{E}{p}
=
\frac{-
\frac{w}{2 X}
\dot{\phi} (\n \cdot \nabla \phi)
\pm \sqrt{
1 
+ 
\frac{w}{2 X}
(\dot{\phi}^{2}
- 
(\n \cdot \nabla \phi)^{2} )
+
(1 + \frac{w \dot{\phi}^{2} }{2 X})
\frac{m^{2}}{p^{2}}
}
}{1 + \frac{w \dot{\phi}^{2} }{2 X}}
\label{Sol1}
\end{gather}
%+++++++++++++++++++++++++++++++++++++++++++++++++++++++++++++++++++++++++++++++
The choice of either plus or minus sign in the solution reflects the two particle/anti-particle states of the neutrino, with the plus sign corresponding to the neutrino and the minus sign corresponding to the anti-neutrino. According to the Feynman-Stueckelberg interpretation of anti-particles, the anti-neutrino can be thought of as a positive energy neutrino traveling backwards in time. Thus, the solution (\ref{Sol1}) with the negative sign, representing the anti-neutrino energy, should have a overall negative sign removed. Furthermore, since time is reversed, this means that in order to have the anti-neutrino traveling in the direction of $\n$, we must replace $\n \rightarrow - \n$ when we choose the negative sign in (\ref{Sol1}). In the end, the energy-momentum relation for the neutrino and anti-neutrino will be the same and given by (\ref{Sol1}) with the positive sign. If we expand (\ref{Sol1}) to first order in $u$ and $m^{2}$, then $\frac{E}{p}$ for the (anti-)neutrino becomes 
%+++++++++++++++++++++++++++++++++++++++++++++++++++++++++++++++++++++++++++++++
\begin{gather}
\frac{E}{p}
\approx
1
-
\frac{g u}{2 X}
\left(
\dot{\phi}
+
\n \cdot \nabla \phi
\right)^{2}
+
\frac{m^{2}}{2 p^{2}}
\label{Sol2}
\end{gather}
%+++++++++++++++++++++++++++++++++++++++++++++++++++++++++++++++++++++++++++++++
From (\ref{Sol1}), we find that the group velocity is
%+++++++++++++++++++++++++++++++++++++++++++++++++++++++++++++++++++++++++++++++
\begin{gather}
v_{g}
=
\frac{1}{1 + \frac{w \dot{\phi}^{2}}{2 X}}
\left(
\frac{1 + \frac{w}{2 X} ( \dot{\phi}^{2} - (\n \cdot \nabla \phi)^{2}) }{\sqrt{1 + \frac{w}{2 X} ( \dot{\phi}^{2} - (\n \cdot \nabla \phi)^{2} ) + (1 + \frac{w \dot{\phi}^{2}}{2 X}) \frac{m^{2}}{p^{2}}}}
-
\frac{w}{2 X}
\dot{\phi}
(\n \cdot \nabla \phi)
\right)
\label{GeneralGroupVelocity}
\end{gather}
%+++++++++++++++++++++++++++++++++++++++++++++++++++++++++++++++++++++++++++++++
It is important to note that the neutrino velocity (\ref{GeneralGroupVelocity}) does not change under a redefinition of the k-essence field variable $\phi$ unless $m^{2} \neq 0$. It is easy to show that under a redefinition from $\phi$ to another field $\varphi$ defined by $\phi = g(\varphi)$, then (\ref{GeneralGroupVelocity}) would remain unchanged were it not for the the $m^{2}$ term in the denominator. Recall then $m^{2}$ is not the physical neutrino mass but rather a rescaled mass, which is rescaled by the conformal factor $\Omega$ in the emergent metric. After a field rescaling the effective neutrino mass becomes
%+++++++++++++++++++++++++++++++++++++++++++++++++++++++++++++++++++++++++++++++
\begin{gather}
m^{2}
=
\frac{M^{2}}{\Omega^{2}}
=
M^{2}
\frac{L_{X}}{c_{s}}
\quad
\rightarrow
\quad
m^{2}
=
\frac{M^{2}}{[g'(\varphi)]^{2}}
\frac{L_{\tilde{X}}}{c_{s}}
=
\frac{\tilde{m}^{2}}{[g'(\varphi)]^{2}}
\end{gather}
%+++++++++++++++++++++++++++++++++++++++++++++++++++++++++++++++++++++++++++++++
Before any objections are raised we should point out that from the beginning we have chosen a specific background k-essence field that the neutrino couples to. In essence what we have done is to fix the ``gauge'' of the k-essence field. Therefore, it is no surprise that by changing the field variable we are changing the physics of the neutrino field. Since a field redefinition changes that conformal factor $\Omega$, a field redefinition is itself a conformal transformation. That the effective mass is the only quantity that changes under a field redefinition is a reflection of the fact that by adding a neutrino mass we are in essence breaking the conformal invariance of the neutrino action. 
%+++++++++++++++++++++++++++++++++++++++++++++++++++++++++++++++++++++++++++++++

%+++++++++++++++++++++++++++++++++++++++++++++++++++++++++++++++++++++++++++++++
%+++++++++++++++++++++++++++++++++++++++++++++++++++++++++++++++++++++++++++++++
%+++++++++++++++++++++++++++++++++++++++++++++++++++++++++++++++++++++++++++++++
%+++++++++++++++++++++++++++++++++++++++++++++++++++++++++++++++++++++++++++++++
\subsection{Comparisons with Observation}
%+++++++++++++++++++++++++++++++++++++++++++++++++++++++++++++++++++++++++++++++
%+++++++++++++++++++++++++++++++++++++++++++++++++++++++++++++++++++++++++++++++
%+++++++++++++++++++++++++++++++++++++++++++++++++++++++++++++++++++++++++++++++
%+++++++++++++++++++++++++++++++++++++++++++++++++++++++++++++++++++++++++++++++
In 1987 a supernova was observed \cite{Hirata:1987hu,Bionta:1987qt} in the Large Magellanic Cloud that provided a limited but unique opportunity for the study of neutrino physics. During this event three neutrinos where detected at neutrino detectors here on earth and unambiguous identified as having been the emitted by the supernova. By comparing the time interval between when the supernova was first seen and when the neutrinos were detected a limit on the neutrino speed's deviation from the speed of light could be bounded. In \cite{Stodolsky:1987vd} the authors calculated, using the available data from the supernova event, that that deviation of the neutrino speed from the speed of light can not be more than 1 part in $10^{8}$. In order words if $v_{\nu}$ is the neutrino speed then 
%+++++++++++++++++++++++++++++++++++++++++++++++++++++++++++++++++++++++++++++++
\begin{gather}
\left|\frac{c}{v_{\nu}} -1 \right|
< 10^{-8}.
\label{vnubound}
\end{gather}
%+++++++++++++++++++++++++++++++++++++++++++++++++++++++++++++++++++++++++++++++
If the neutrino is massless then $v_{\nu} = c_{\nu}$, and the left hand side of (\ref{vnubound}) is equal to $|gu|$ in the physically relevant static case\footnote{note that $c=1$ in all previous sections}. Thus, in the massless neutrino limit (\ref{vnubound}) represents an observationally required upper bound on $|g u|$. In the most general case this bound becomes (setting $c=1$ again)
%+++++++++++++++++++++++++++++++++++++++++++++++++++++++++++++++++++++++++++++++
\begin{gather}
\left|\frac{1}{v_{\nu}} -1 \right|
< 10^{-8}
\quad
\rightarrow
\quad
\left|
\frac{g u}{2 X}
(\dot{\phi} + \n \cdot \nabla \phi)^{2}
+
\frac{m^{2}}{2 E^{2}}
\right|
<10^{-8}
\end{gather}
%++++++++++++++++ +++++++++++++++++++++++++++++++++++++++++++++++++++++++++++++++
The most generous upper bound that can be placed on $g u$ is $\left| g u \right| < 10^{-8}$. This can also be translated into a restriction on the k-essence sound speed $c_{s}$. If the k-essence field has a time like gradient (i.e. $X>0$) then the range of values for $c_{s}$ is 
%+++++++++++++++++++++++++++++++++++++++++++++++++++++++++++++++++++++++++++++++
\begin{gather}
\frac{1}{1+  10^{-8} / |g|} 
< 
c_{s}
<
\frac{1}{1- 10^{-8}/|g|} 
\label{UniformCSConstraint}
\end{gather}
%+++++++++++++++++++++++++++++++++++++++++++++++++++++++++++++++++++++++++++++++
%In the static case however, $u= c_{s} -1$ and the bound on $c_{s}$ is
%+++++++++++++++++++++++++++++++++++++++++++++++++++++++++++++++++++++++++++++++
%\begin{gather}
%1 -  10^{-8} / |g|
%< 
%c_{s}
%<
%1 +  10^{-8}/|g|
%\label{StaticCSConstraint}
%\end{gather}
%+++++++++++++++++++++++++++++++++++++++++++++++++++++++++++++++++++++++++++++++
%If $10^{-8} /|g| \ll 1$ then both (\ref{UniformCSConstraint}) reduces to (\ref{StaticCSConstraint}) and we have a model independent restriction on $c_{s}$. 
As of this moment there is an insufficient amount of data to constrain $c_{s}$ from cosmological observables. Recent studies have shown that fits of general dark energy models to the current CMBR data are largely insensitive to the value of $c_{s}$ \cite{Hannestad:2005ak,Xia:2007km}. Without any other observational constraints on (or better yet a value for) $c_{s}$, it is impossible to reliably estimate the value of the k-essence coupling $g$. However, with more precise data in the future it may be possible to get a better handle on the value of $c_{s}$, and once that has been established, (\ref{UniformCSConstraint}) can be used to put useful restrictions on $g$.
%+++++++++++++++++++++++++++++++++++++++++++++++++++++++++++++++++++++++++++++++

%+++++++++++++++++++++++++++++++++++++++++++++++++++++++++++++++++++++++++++++++
%+++++++++++++++++++++++++++++++++++++++++++++++++++++++++++++++++++++++++++++++
%+++++++++++++++++++++++++++++++++++++++++++++++++++++++++++++++++++++++++++++++
%+++++++++++++++++++++++++++++++++++++++++++++++++++++++++++++++++++++++++++++++
\section{Neutrino Oscillations}
\label{NeutrinoOscillations}
%+++++++++++++++++++++++++++++++++++++++++++++++++++++++++++++++++++++++++++++++
Experiments \cite{Fukuda:1998mi} have confirmed the the phenomenon of flavor oscillations whereby neutrinos oscillate between various possible flavor eigenstates as they travel away from their source. There are different ways of explaining this oscillation, but all mechanisms for inducing neutrino oscillations involve some term in the neutrino lagrangian that is non-diagonal in the flavor eigenstates. Although the most popular way for inducing neutrino oscillations is by introducing a mass term \cite{Pontecorvo:1957qd}, several other mechanism have been proposed over the years, such as: violation of the equivalence principle (VEP) \cite{Gasperini:1988zf,Halprin:1991gs}, torsion induced neutrino oscillations \cite{DeSabbata:1981ek}, violation of Lorentz invariance (VLI) \cite{Coleman:1997xq}, violation of CPT symmetry \cite{Coleman:1998ti}. 
%+++++++++++++++++++++++++++++++++++++++++++++++++++++++++++++++++++++++++++++++

%+++++++++++++++++++++++++++++++++++++++++++++++++++++++++++++++++++++++++++++++
If neutrinos do indeed couple to a k-essence background in the manner we described in \S\ref{BackGround}, then it is possible that k-essence can play a role in neutrino oscillations. There are two ways in which k-essence could effect neutrino oscillations. If the k-essence coupling is the same for each neutrino flavor then the energy difference between energy eigenstates will not be affected. However, neutrinos coupled to the k-essence background will travel along geodesics in the emergent space-time. This will have an affect on the phase of the neutrino wavefunction, which will be observable in the neutrino oscillation probability \cite{Cardall:1996cd,Papini:2001kb,Fornengo:1997qu,Kojima:1996vb}.
%+++++++++++++++++++++++++++++++++++++++++++++++++++++++++++++++++++++++++++++++

%+++++++++++++++++++++++++++++++++++++++++++++++++++++++++++++++++++++++++++++++
Another way k-essence can influence neutrino oscillations is if the k-essence coupling $g$ is non-diagonal in the flavor eigenstate basis. Imagine now a model of two neutrino flavors that couple non-diagonally to k-essence in the flavor eigenbasis. The lagrangian of this 2-neutrino system can be written as ($\alpha, \beta =1,2$ are flavor indices),
%+++++++++++++++++++++++++++++++++++++++++++++++++++++++++++++++++++++++++++++++
\begin{gather}
\L
=
i
\sum_{\alpha=1,2}
E_{\alpha}
\bar{\nu}_{K \alpha}
\tilde{\gamma}_{\alpha}^{\mu}
\nabla_{\mu}
\nu_{K \alpha}
-
\frac{1}{2}
\sum_{\alpha,\beta = 1,2}
\bar{\nu}_{K \alpha}
\left(
M_{\alpha\beta}
E_{\beta}
+
E_{\alpha}
M_{\alpha \beta}
\right)
\nu_{K \beta}
\end{gather}
%+++++++++++++++++++++++++++++++++++++++++++++++++++++++++++++++++++++++++++++++
where $E_{\alpha} = E_{\alpha \alpha}$ and $E_{\alpha \beta} =[\det E_{\mu}^{a}(\hat{g})]_{\alpha \beta}$. In this lagrangian the k-essence coupling $g$ has been replaced by a matrix valued object $\hat{g}$, that is not necessarily diagonal in the flavor and mass eigenstates. We have defined $\nu_{K}$ as the ``k-essence eigenstates'', which are the eigenstates of the k-essence coupling matrix $\hat{g}$. In general the k-essence eigenstates will not be the same as the neutrino flavor eigenstates. Because of this non-diagonal coupling of k-essence to the flavor eigenstates, the formula for the oscillation probability in the case of k-essence induced neutrino oscillations (KINO) will differ significantly from the typical mass-induced result.
%+++++++++++++++++++++++++++++++++++++++++++++++++++++++++++++++++++++++++++++++

%+++++++++++++++++++++++++++++++++++++++++++++++++++++++++++++++++++++++++++++++
In this section we will assume that the k-essence field is weakly varying, with small second derivatives so that we may effectively treat the k-essence interaction, which only involves the first derivatives of the scalar field, as a coupling constant. This greatly simplifies matters, because it allows us to diagonalize the neutrino equations of motion in the momentum space representation. This assumption is  consistent with the literature where most models of k-essence take the field and its sound speed to be relatively constant in time and space. As a result of this assumption we can absorb the determinants $E_{\alpha}$ into a redefinition of the neutrino wavefunction $\nu_{K}$ and mass matrix $M_{\alpha\beta}$. Therefore, it is safe to ignore this contribution in our analysis. 

In the first part of this section we will consider the effect of k-essence on neutrino oscillations when the k-essence coupling is equal for each flavor eigenstate. 
%+++++++++++++++++++++++++++++++++++++++++++++++++++++++++++++++++++++++++++++++

%+++++++++++++++++++++++++++++++++++++++++++++++++++++++++++++++++++++++++++++++
%+++++++++++++++++++++++++++++++++++++++++++++++++++++++++++++++++++++++++++++++
%+++++++++++++++++++++++++++++++++++++++++++++++++++++++++++++++++++++++++++++++
%+++++++++++++++++++++++++++++++++++++++++++++++++++++++++++++++++++++++++++++++
%+++++++++++++++++++++++++++++++++++++++++++++++++++++++++++++++++++++++++++++++
%+++++++++++++++++++++++++++++++++++++++++++++++++++++++++++++++++++++++++++++++
%+++++++++++++++++++++++++++++++++++++++++++++++++++++++++++++++++++++++++++++++
\subsection{Neutrino Oscillations with Flavor Diagonal K-essence Couplings}
\label{DiagCase}
%+++++++++++++++++++++++++++++++++++++++++++++++++++++++++++++++++++++++++++++++
%+++++++++++++++++++++++++++++++++++++++++++++++++++++++++++++++++++++++++++++++
%+++++++++++++++++++++++++++++++++++++++++++++++++++++++++++++++++++++++++++++++
%+++++++++++++++++++++++++++++++++++++++++++++++++++++++++++++++++++++++++++++++
%+++++++++++++++++++++++++++++++++++++++++++++++++++++++++++++++++++++++++++++++
At first it may seem that when both neutrino flavors couple to k-essence identically, the usual formula for mass -induced neutrino oscillations should not change. However, even in this case the fact that neutrinos travel on geodesics in the emergent space-time will create deviations from the flat space result. In order to best analyze neutrino oscillations in the presence of a K-essence background, we will use the simple method used in \cite{Cardall:1996cd,Piriz:1996mu} to study neutrino oscillations in curved space-times. In all cases, the important quantity of interest when calculating the neutrino oscillation probability is the phase of the neutrino wavefunction.  
%+++++++++++++++++++++++++++++++++++++++++++++++++++++++++++++++++++++++++++++++
\begin{gather}
\left|\nu_{f}(t,x)
\right>
=
e^{- i \hat{\Phi}}
\left|
\nu_{f}
\right>
=
e^{- i \int \hat{p}_{\mu} dx^{\mu}}
\left|
\nu_{f}
\right>
\label{NeutrinoPhase}
\end{gather}
%+++++++++++++++++++++++++++++++++++++++++++++++++++++++++++++++++++++++++++++++
The difference between the phases $\hat{\Phi}$ of two different flavors of neutrino will lead to neutrino oscillations. In the expression (\ref{NeutrinoPhase}), the hats over $\hat{\Phi}$ and the 4-momentum $\hat{p}_{\mu}$ indicate that these are operators which act in the flavor space of neutrinos. The phase $\hat{\Phi}$ when written out in terms of the energy and three momentum is 
%+++++++++++++++++++++++++++++++++++++++++++++++++++++++++++++++++++++++++++++++
\begin{gather}
\hat{\Phi}
=
\int 
E
\hat{I}
dt
-
\hat{\p} \cdot d\x
\label{NeutrinoPhaseOpTwo}
\end{gather}
%+++++++++++++++++++++++++++++++++++++++++++++++++++++++++++++++++++++++++++++++
where $\hat{I}$ is the identity operator\footnote{Note, we could have assumed that the momentum operator was proportional to the identity and the energy $E$ was an off diagonal  operator. Both approaches are equivalent when a first order expansion in the energy has been taken, as it will be in the final answer.}. In a full and proper treatment of neutrino oscillations, the neutrinos must be modeled by spatially localized wave packets composed of neutrino energy eigenstates. However, if one uses this approach to study neutrino oscillations, they would find that in the relativistic limit it is acceptable to assume that both neutrino eigenstates propagate on the same null geodesic between the neutrino emitter and detector in space-time \cite{Kayser:1981ye,Rich:1993wu,Giunti:1993se}\footnote{In the wavepacket treatment, an exponential damping is found in the final result for the oscillation probability in the relativistic limit. This is due to the decoherence of the superposition of neutrino wavefunctions. So far there has been no solid evidence for decoherence effects \cite{Fogli:2003th}, and we are therefore safe in ignoring this possibility in other analysis}. Therefore the phase operator (\ref{NeutrinoPhaseOpTwo}) becomes
%+++++++++++++++++++++++++++++++++++++++++++++++++++++++++++++++++++++++++++++++
\begin{gather}
\hat{\Phi}
=
\int_{t_{e}}^{t_{d}}
\left(
E
\hat{I}
-
\hat{\p}
\cdot
\left[\frac{d\x}{dt}\right]_{0}
\right)
dt
\label{PhaseTwo}
\end{gather}
%+++++++++++++++++++++++++++++++++++++++++++++++++++++++++++++++++++++++++++++++
where $t_{e}$ and $t_{d}$ are the values of the coordinate time at which the neutrino signal is emitted and detected. The ``0'' subscript on $\frac{d\x}{dt}$ denotes that this quantity is to be evaluated along a null geodesic between the emitter and detector. To find $\frac{d\x}{dt}$, we start from the definition of the canonical momentum of a massive neutrino 
%+++++++++++++++++++++++++++++++++++++++++++++++++++++++++++++++++++++++++++++++
\begin{gather}
p_{\mu}
=
m G_{\mu\nu}
\frac{d x^{\nu}}{d s}
\end{gather}
%+++++++++++++++++++++++++++++++++++++++++++++++++++++++++++++++++++++++++++++++
Here $s$ is a proper time coordinate defined in the neutrino rest frame. With this, it is clear that $\frac{d\x}{dt}$ is
%+++++++++++++++++++++++++++++++++++++++++++++++++++++++++++++++++++++++++++++++
\begin{gather}
\frac{d\x}{dt}
=
\frac{\frac{d\x}{ds}}{\frac{dt}{ds}}
=
\frac{G^{i\mu}p_{\mu}}{G^{0 \nu} p_{\nu}}
=
\frac{\p - \frac{w}{2 X} (\dot{\phi} E + \p \cdot \nabla \phi) \nabla \phi}{E + \frac{w}{2 X} (\dot{\phi} E + \p \cdot \nabla \phi) \dot{\phi}}
\end{gather}
%+++++++++++++++++++++++++++++++++++++++++++++++++++++++++++++++++++++++++++++++
Along a null geodesic the energy and momentum of a massless neutrino are denoted by $E^{(0)}$ and $p^{(0)}$. Let us denote the unit vector in the direction of the neutrino momentum by $\n$. Thus, $\n \cdot \left[\frac{d\x}{dt} \right]_{0}$ is given by
%+++++++++++++++++++++++++++++++++++++++++++++++++++++++++++++++++++++++++++++++
\begin{gather}
\n \cdot \left[\frac{d\x}{dt} \right]_{0}
=
\frac{- \frac{w}{2 X}\dot{\phi} (\n \cdot \nabla \phi)+ \sqrt{1 + \frac{w}{2 X} (\dot{\phi}^{2} - (\n \cdot \nabla \phi)^{2})}}{1 + \frac{w \dot{\phi}^{2}}{2 X}}
=
\frac{E^{(0)}}{p^{(0)}}
\label{dxdt0}
\end{gather}
%+++++++++++++++++++++++++++++++++++++++++++++++++++++++++++++++++++++++++++++++
To relate $E$ and $\p$, recall the discussion from section \S\ref{NonStaticNonUniformVelocity}. It is easy to show from the work done there that the momentum as a function of $E$ expressed to leading order in $m^{2}$ is
%+++++++++++++++++++++++++++++++++++++++++++++++++++++++++++++++++++++++++++++++
\begin{gather}
p
\approx
E\frac{\frac{w}{2 X} \dot{\phi} (\n \cdot \nabla \phi) + \sqrt{1 + \frac{w}{2 X} (\dot{\phi}^{2} - (\n \cdot \nabla \phi)^{2})}}{1 - \frac{w}{2 X} (\n \cdot \nabla \phi)^{2}}
-
\frac{\frac{m^{2}}{2 E}}{\sqrt{1 + \frac{w}{2 X} (\dot{\phi}^{2} - (\n \cdot \nabla \phi)^{2})}}
\label{PasfuncofE}
\end{gather}
%+++++++++++++++++++++++++++++++++++++++++++++++++++++++++++++++++++++++++++++++
Now remember that $m^{2} = \Omega^{-2} M^{2}$. In order for there to be neutrino oscillations the flavor and mass eigenstates must not be the same. Therefore, we have to replace the $M^{2}$ with a matrix $\hat{M}^{2}$ that is non-diagonal in the flavor eigenstate basis. Since only phase differences between flavor eigenstates will be important, we can ignore terms not proportional to $\hat{M}^{2}$ in the phase operator (\ref{PhaseTwo}). With (\ref{dxdt0}) and (\ref{PasfuncofE}) the expression for the phase operator (\ref{PhaseTwo}), modulo terms proportional to the identity,  becomes
%+++++++++++++++++++++++++++++++++++++++++++++++++++++++++++++++++++++++++++++++
\begin{gather}
\hat{\Phi}
=
\frac{\hat{M}^{2} L(\ell)}{2 E^{(0)}}
\label{PhaseOpOne}
\end{gather}
%+++++++++++++++++++++++++++++++++++++++++++++++++++++++++++++++++++++++++++++++
where $x_{e}$ and $x_{d}$ are the neutrino and detector positions, respectively. We have defined an effective distance $L$, which is a function of the coordinate distance $\ell = |x_{d} - x_{e}|$ between the neutrino emitter and detector. The expression for $L(\ell)$ is given by
%+++++++++++++++++++++++++++++++++++++++++++++++++++++++++++++++++++++++++++++++
\begin{gather}
L(\ell)
=
\int_{x_{e}}^{x_{d}}
\frac{\Omega^{-2}(\n \cdot d\x)}{\sqrt{1 + \frac{w}{2 X} (\dot{\phi}^{2} - (\n \cdot \nabla \phi)^{2})}}
\end{gather}
%+++++++++++++++++++++++++++++++++++++++++++++++++++++++++++++++++++++++++++++++
In general this distance will differ from the true emitter/detector separation due to the neutrino coupling to the k-essence medium. There are two sets of neutrino eigenstates in this system: the flavor eigenstates $\left|\nu_{f}\right> = \{\left|\nu_{e}\right>, \left|\nu_{\mu}\right>\}$ which couple diagonally to the weak current, and the mass eigenstates $\left|\nu_{m}\right> = \{\left|\nu_{m_{1}}\right>, \left|\nu_{m_{2}}\right>\}$ that define the basis in which the mass matrix is diagonal. In general these two sets will not be equivalent. However, we can define an $SU(2)$ matrix $V$ such that 
%+++++++++++++++++++++++++++++++++++++++++++++++++++++++++++++++++++++++++++++++
\begin{gather}
\left|
\nu_{f}
\right>
=
\sum_{m}
V_{fm}
\left|
\nu_{m}
\right>
\end{gather}
%+++++++++++++++++++++++++++++++++++++++++++++++++++++++++++++++++++++++++++++++
The matrix $V$ is referred to as the mass mixing matrix. The most general $SU(2)$ matrix can be represented in terms of an angle and three phases:
%+++++++++++++++++++++++++++++++++++++++++++++++++++++++++++++++++++++++++++++++
\begin{gather}
e^{i \chi}
\left[
\begin{array}{cc}
e^{- i \alpha} & 0
\\
0 & e^{i \alpha}
\end{array}
\right]
\left[
\begin{array}{cc}
\cos\theta & \sin\theta
\\
-\sin\theta & \cos\theta
\end{array}
\right]
\left[
\begin{array}{cc}
e^{- i \beta} & 0
\\
0 & e^{i \beta}
\end{array}
\right]
\end{gather}
%+++++++++++++++++++++++++++++++++++++++++++++++++++++++++++++++++++++++++++++++
In general, $V$ is an $SU(2)$ matrix and will therefore have three additional phase degrees of freedom. However, in this case we can absorb these phases into a redefinition of the neutrino wavefunction. Thus we can safely ignore these phases in $V$ and as a result the mixing between the flavor and mass eigenstates is determined by a single angle which we will call $\theta_{M}$:
%+++++++++++++++++++++++++++++++++++++++++++++++++++++++++++++++++++++++++++++++
\begin{gather}
V
=
\left[
\begin{array}{cc}
\cos \theta_{M} & \sin \theta_{M}
\\
- \sin \theta_{M} & \cos \theta_{M}
\end{array}
\right]
\end{gather}
%+++++++++++++++++++++++++++++++++++++++++++++++++++++++++++++++++++++++++++++++
Written in the flavor eigenstate basis the phase operator (\ref{PhaseOpOne}) is 
%+++++++++++++++++++++++++++++++++++++++++++++++++++++++++++++++++++++++++++++++
\begin{gather}
{\footnotesize
\hat{\Phi}(\ell)
=
\frac{ L(\ell)}{2 E}
V
\left[
\begin{array}{cc}
m_{1}^{2} & 0
\\
0 & m_{2}^{2}
\end{array}
\right]
V^{\dagger}
=
\frac{\Delta m^{2} L(\ell)}{4 E}
\left[
\begin{array}{cc}
-\cos 2 \theta_{M} & \sin 2 \theta_{M}
\\
\sin 2 \theta_{M} &  \cos 2 \theta_{M}
\end{array}
\right]
+
\frac{\bar{m}^{2} L(\ell)}{2 E}
\left[
\begin{array}{cc}
1 & 0
\\
0 & 1
\end{array}
\right]}
\label{PhaseMatrix}
\end{gather}
%+++++++++++++++++++++++++++++++++++++++++++++++++++++++++++++++++++++++++++++++
where $m_{1}$ and $m_{2}$ are the masses of the two neutrino mass eigenstates, and $\Delta m^{2} = m_{2}^{2} - m_{1}^{2}$ and $\bar{m}^{2} = \frac{m_{2}^{2} + m_{1}^{2}}{2}$.  Note that we have dropped the ``$(0)$'' superscript on the energy $E$ merely out of convenience. Since we are interested in finding probabilities then we can subtract from the phase matrix any term proportional to the identity matrix without changing our final result. Therefore, we will ignore the very last term in (\ref{PhaseMatrix}) from here on out, and the phase factor in (\ref{NeutrinoPhase}) becomes
%+++++++++++++++++++++++++++++++++++++++++++++++++++++++++++++++++++++++++++++++
\begin{gather}
e^{- i \hat{\Phi}}
=
-i
\sin(\frac{\Delta m^{2} L(\ell) }{4 E})
\left[
\begin{array}{cc}
-\cos 2 \theta_{M} & \sin 2 \theta_{M}
\\
\sin 2 \theta_{M} &  \cos 2 \theta_{M}
\end{array}
\right]
+
\cos(\frac{\Delta m^{2} L(\ell) }{4 E})
\left[
\begin{array}{cc}
1 & 0
\\
0 & 1
\end{array}
\right]
\label{evoop}
\end{gather}
%+++++++++++++++++++++++++++++++++++++++++++++++++++++++++++++++++++++++++++++++
The factor $e^{- i \hat{\Phi}}$ plays the role of a evolution operator. Denote by $\left|\nu_{f}\right>$ the initial state of a neutrino of flavor $f$. The time at which the neutrino is created can be defined, without loss of generality, as occurring at $t=0$. The wavefunction of this neutrino at some later time $t > 0$ after it has traveled a distance $\ell$ away from its source will be given by the action of $e^{- i \hat{\Phi}(\ell)}$ on the neutrino wavefunction $\left|\nu_{f}\right>$:
%+++++++++++++++++++++++++++++++++++++++++++++++++++++++++++++++++++++++++++++++
\begin{gather}
\left|
\nu_{f}(\ell)
\right>
=
\sum_{f'}
[e^{- i \hat{\Phi}(\ell)}]_{ff'}
\left|
\nu_{f'}
\right>
\end{gather}
%+++++++++++++++++++++++++++++++++++++++++++++++++++++++++++++++++++++++++++++++
Applying the phase operator (\ref{evoop}) to the neutrino ket $\left|\nu_{f} \right>$, we find that the flavor eigenstate wavefunctions at a later time $t>0$ when the neutrinos have traveled a distance $\ell$ are
%+++++++++++++++++++++++++++++++++++++++++++++++++++++++++++++++++++++++++++++++
\begin{gather}
\left|\nu_{e}(\ell)\right>
=
i 
\sin (\frac{\Delta m^{2} L(\ell) }{4 E})
\left(
\cos 2 \theta_{M}
\left|
\nu_{e}
\right>
-
\sin 2 \theta_{M}
\left|
\nu_{\mu}
\right>
\right)
+
\cos (\frac{\Delta m^{2} L(\ell) }{4 E})
\left|
\nu_{e}
\right>
\\
\left|\nu_{\mu}(\ell)\right>
=
-i
\sin (\frac{\Delta m^{2} L(\ell) }{4 E})
\left(
\sin 2 \theta_{M}
\left|
\nu_{e}
\right>
+
\cos 2 \theta_{M}
\left|
\nu_{\mu}
\right>
\right)
+
\cos
(\frac{\Delta m^{2} L(\ell) }{4 E})
\left|
\nu_{\mu}
\right>
\end{gather}
%+++++++++++++++++++++++++++++++++++++++++++++++++++++++++++++++++++++++++++++++
Suppose a neutrino of flavor $f$ is created from some source of interest (e.g. the sun, atmosphere, nuclear reactor, etc...) at some time $t_{e}$. The probability of this neutrino appearing as an $f'$ flavored neutrino in a detector here on earth at some later time $t_{d}$ after having traversed a distance $\ell$ is given by the expression 
%+++++++++++++++++++++++++++++++++++++++++++++++++++++++++++++++++++++++++++++++
\begin{gather}
P(\nu_{f} \rightarrow \nu_{f'})
=
|\left<
\nu_{f'}
\right|
\left.
\nu_{f}(\ell)
\right>|^{2}
=
\delta_{f,f'}
+
(-1)^{\delta_{f,f'}}
\sin^{2}2 \theta_{M}
\sin^{2} ( \frac{\Delta m^{2} L(\ell) }{4 E} )
\end{gather}
%+++++++++++++++++++++++++++++++++++++++++++++++++++++++++++++++++++++++++++++++
As we can see here the formula for the oscillation probability is almost exactly the same as in the standard mass-induced result. The only difference is the effective distance $L$ is used instead of the coordinate distance $\ell$. 
%+++++++++++++++++++++++++++++++++++++++++++++++++++++++++++++++++++++++++++++++

%+++++++++++++++++++++++++++++++++++++++++++++++++++++++++++++++++++++++++++++++
%+++++++++++++++++++++++++++++++++++++++++++++++++++++++++++++++++++++++++++++++
%+++++++++++++++++++++++++++++++++++++++++++++++++++++++++++++++++++++++++++++++
%+++++++++++++++++++++++++++++++++++++++++++++++++++++++++++++++++++++++++++++++
%+++++++++++++++++++++++++++++++++++++++++++++++++++++++++++++++++++++++++++++++
%+++++++++++++++++++++++++++++++++++++++++++++++++++++++++++++++++++++++++++++++
%+++++++++++++++++++++++++++++++++++++++++++++++++++++++++++++++++++++++++++++++
\subsection{Neutrino Oscillations with Flavor Non-diagonal K-essence Couplings}
\label{NonDiagCase}
%+++++++++++++++++++++++++++++++++++++++++++++++++++++++++++++++++++++++++++++++
%+++++++++++++++++++++++++++++++++++++++++++++++++++++++++++++++++++++++++++++++
%+++++++++++++++++++++++++++++++++++++++++++++++++++++++++++++++++++++++++++++++
%+++++++++++++++++++++++++++++++++++++++++++++++++++++++++++++++++++++++++++++++
%+++++++++++++++++++++++++++++++++++++++++++++++++++++++++++++++++++++++++++++++
At the beginning on this paper we have introduced a k-essence/neutrino coupling parameter that we have denoted by $g$. In the last subsection we assumed that this coupling was the same for each neutrino flavor. In this subsection, however, we will consider what happens when this coupling $g$ becomes a matrix valued object $\hat{g}$ that operates in the neutrino flavor vector space. In this case the equation for the energy/momentum relation (\ref{Sol2}) becomes a matrix equation whose eigenvalues represent that energies of the neutrino energy eigenstates which in general are not the same as the flavor, mass and k-essence eigenstates. A formula for the momentum as a function of energy can be derived from the information found in section \S\ref{NonStaticNonUniformVelocity}. If we expand this formula around $g=0$ and $m^{2} = 0$ to leading order, and then replace $g$ and $M^{2} = \Omega^{2} m^{2}$ with matrices that act in the neutrino flavor space we end up with
%+++++++++++++++++++++++++++++++++++++++++++++++++++++++++++++++++++++++++++++++
\begin{gather}
\hat{p}
\approx
E
\hat{I}
+
\frac{E \hat{G}}{2}
-
\frac{\hat{M}^{2} \Omega^{-2}}{2 E}
\end{gather}
%+++++++++++++++++++++++++++++++++++++++++++++++++++++++++++++++++++++++++++++++
where $\hat{G} = \frac{\hat{g} u}{X} (\dot{\phi} + \n \cdot \nabla \phi)^{2}$. Here $\hat{G}$ and $\hat{M}^{2}$ are operators in the flavor space. In general it will not be possible to diagonalize $\hat{G}$ and $\hat{M}^{2}$ at the same time, since there is no reason to assume that the k-essence and mass eigenstates are the same. We can relate the flavor, mass, and k-essence eigenstates by two $SU(2)$ matrices. Define two $SU(2)$ matrices $U_{f\alpha}$ and $V_{fm}$ such that
%+++++++++++++++++++++++++++++++++++++++++++++++++++++++++++++++++++++++++++++++
\begin{gather}
\left|\nu_{f}\right>
=
\sum_{m}
V_{fm}
\left|
\nu_{m}
\right>
,
\quad
\left|\nu_{f}\right>
=
\sum_{\alpha}
U_{f\alpha}
\left|
\nu_{\alpha}
\right>
\quad
\Rightarrow
\quad
\left|
\nu_{\alpha}
\right>
=
\sum_{f,m}
U^{\dagger}_{f\alpha}
V_{f m}
\left|\nu_{m}\right>
\end{gather}
%+++++++++++++++++++++++++++++++++++++++++++++++++++++++++++++++++++++++++++++++
In this case where there are three sets of neutrino eigenstates, one can not simply disregard the phases in the k-essence and mass mixing matrices $U_{f\alpha}$ and $V_{f m}$ as can be done in the standard treatment of purely mass-induced neutrino oscillations. While we can eliminate most of the phases through a redefinition of the neutrino eigenstates, there is not enough freedom to get rid of them all. After any redefinition of the neutrino eigenstates, there will still be a single overall phase left. Let's call this residual phase $\alpha$ and define the k-essence and mass mixing matrices as
%+++++++++++++++++++++++++++++++++++++++++++++++++++++++++++++++++++++++++++++++
\begin{gather}
U
=
\left[
\begin{array}{cc}
e^{- i \alpha}\cos\theta_{K} & e^{- i \alpha}\sin\theta_{K}
\\
-e^{i \alpha} \sin\theta_{K} & e^{i \alpha}\cos\theta_{K}
\end{array}
\right]
,
\qquad
V
=
\left[
\begin{array}{cc}
\cos\theta_{M} & \sin\theta_{M}
\\
-\sin\theta_{M} & \cos\theta_{M}
\end{array}
\right]
\end{gather}
%+++++++++++++++++++++++++++++++++++++++++++++++++++++++++++++++++++++++++++++++
If $G_{1}$ and $G_{2}$ are the eigenvalues of $\hat{G}$, the momentum operator can be written as
%+++++++++++++++++++++++++++++++++++++++++++++++++++++++++++++++++++++++++++++++
\begin{gather}
\hat{p}
\approx
E
\left[
\begin{array}{cc}
1 & 0 \\
0 & 1
\end{array}
\right]
+
\frac{E}{2}
U
\left[
\begin{array}{cc}
G_{1} & 0 \\
0 & G_{2}
\end{array}
\right]
U^{\dagger}
-
\frac{1}{2 E}
V
\left[
\begin{array}{cc}
\tilde{m}_{1}^{2} & 0 \\
0 & \tilde{m}_{2}^{2}
\end{array}
\right]
V^{\dagger}
\end{gather}
%+++++++++++++++++++++++++++++++++++++++++++++++++++++++++++++++++++++++++++++++
where $\tilde{m}_{i}^{2} = \Omega^{-2} m_{i}^{2}$. If we substitute this for the momentum operator in the integrand in (\ref{PhaseTwo}) we find that modulo terms proportional to the identity matrix we get  
%+++++++++++++++++++++++++++++++++++++++++++++++++++++++++++++++++++++++++++++++
\begin{gather}
E
\hat{I}
-
\hat{\p}
\cdot
\left[
\frac{d\x}{dt}
\right]_{0}
=
\frac{\Delta \tilde{m}^{2}}{4 E}
\left[
\begin{array}{cc}
 - \cos 2 \theta_{M}
&
 \sin 2 \theta_{M}
\\
\sin 2 \theta_{M}
&
\cos 2 \theta_{M}
\end{array}
\right]
-
\frac{E \Delta G}{4}
\left[
\begin{array}{cc}
 - \cos 2 \theta_{K}
&
 e^{- 2 i \alpha} \sin 2 \theta_{K}
\\
 e^{2 i \alpha} \sin 2 \theta_{K}
&
\cos 2 \theta_{K}
\end{array}
\right]
\nn \\
=
\frac{\Delta \tilde{m}^{2}}{4 E}
\left[
\begin{array}{cc}
-\cos 2 \theta_{M} - y \cos 2 \theta_{K} & \sin 2 \theta_{M} + y \sin 2 \theta_{K} e^{- 2 i \alpha}
\\
\sin 2 \theta_{M} + y \sin 2 \theta_{K} e^{2 i \alpha} & \cos 2 \theta_{M} + y \cos 2 \theta_{K}
\end{array}
\right]
\label{EnergyMatrixEqu}
\end{gather}
%+++++++++++++++++++++++++++++++++++++++++++++++++++++++++++++++++++++++++++++++
where $\Delta \tilde{m}^{2} = \tilde{m}_{2}^{2} - \tilde{m}_{1}^{2}$. Here we have introduced a new parameter $y$, which is defined as $y = - \frac{E^{2}\Delta G}{\Delta \tilde{m}^{2}}$ where $\Delta G = G_{2} - G_{1}$. This operator can be written in a much more compact form by defining new variables:
%+++++++++++++++++++++++++++++++++++++++++++++++++++++++++++++++++++++++++++++++
\begin{gather}
\frac{\Delta M^{2}}{4 E}
\cos 2 \theta_{L}
=
\frac{\Delta \tilde{m}^{2}}{4 E}
\left(
\cos 2 \theta_{M}
+
y
\cos2 \theta_{K}
\right)
\nn \\
\frac{\Delta M^{2}}{4 E}
e^{- i \sigma}
\sin 2 \theta_{L}
=
\frac{\Delta \tilde{m}^{2}}{4 E}
\left(
\sin 2 \theta_{M}
+
y
\sin2 \theta_{K}
e^{- 2 i \alpha}
\right)
\end{gather}
%+++++++++++++++++++++++++++++++++++++++++++++++++++++++++++++++++++++++++++++++
Note that the phase $\sigma$, while present here, will not effect our final result. Inverting these
%+++++++++++++++++++++++++++++++++++++++++++++++++++++++++++++++++++++++++++++++
\begin{gather}
\frac{\Delta M^{2}}{4 E}
=
\frac{|\Delta  \tilde{m}^{2} |}{4 E}
\sqrt{
1
+
2 y
\cos 2 \Theta
+
y^{2}
}
\nn \\
\sin^{2}2 \theta_{L}
=
\frac{\sin^{2}2 \theta_{M}  + 2 y \sin 2 \theta_{M} \sin 2 \theta_{K} \cos 2 \alpha + y^{2} \sin^{2} 2 \theta_{K}}{1 + 2 y \cos 2 \Theta + y^{2}}
\nn \\
\tan\sigma
=
\frac{\sin 2 \theta_{K} \sin 2 \alpha}{
\sin 2 \theta_{M}
+
y
\sin 2 \theta_{K} \cos 2 \alpha}
\end{gather}
%+++++++++++++++++++++++++++++++++++++++++++++++++++++++++++++++++++++++++++++++
where
%+++++++++++++++++++++++++++++++++++++++++++++++++++++++++++++++++++++++++++++++
\begin{gather}
\cos 2 \Theta
=
\cos 2 \theta_{M}
\cos 2 \theta_{K}
+
\sin 2 \theta_{M}
\sin 2 \theta_{K}
\cos 2 \alpha
\end{gather}
%+++++++++++++++++++++++++++++++++++++++++++++++++++++++++++++++++++++++++++++++
With these, equation (\ref{EnergyMatrixEqu}) can be written as
%+++++++++++++++++++++++++++++++++++++++++++++++++++++++++++++++++++++++++++++++
\begin{gather}
E
\hat{I}
-
\hat{\p} 
\cdot 
\left[
\frac{d\x}{dt}
\right]_{0}
=
\frac{\Delta M^{2}}{4 E}
\left[
\begin{array}{cc}
 - \cos 2 \theta_{L}
&
e^{- i \sigma}
\sin 2 \theta_{L}
\\
e^{i \sigma}
\sin 2 \theta_{L}
&
\cos 2 \theta_{L}
\end{array}
\right]
\label{2ndMatrixEqu}
\end{gather}
%+++++++++++++++++++++++++++++++++++++++++++++++++++++++++++++++++++++++++++++++
The phase operator is therefore
%+++++++++++++++++++++++++++++++++++++++++++++++++++++++++++++++++++++++++++++++
\begin{gather}
e^{- i \hat{\Phi}(\ell)}
=
- 
i
\sin \varphi(\ell)
\left[
\begin{array}{cc}
- \cos 2 \theta_{L}  & e^{- i \sigma}  \sin 2 \theta_{L} 
\\
e^{i \sigma} \sin 2 \theta_{L} & \cos 2 \theta_{L}
\end{array}
\right]
+
\cos \varphi(\ell)
\left[
\begin{array}{cc}
1 & 0
\\
0 & 1
\end{array}
\right]
\label{evoop2}
\end{gather}
%+++++++++++++++++++++++++++++++++++++++++++++++++++++++++++++++++++++++++++++++
where $\varphi(\ell) = \int_{x_{e}}^{x_{d}} \frac{\Delta M^{2}}{4 E} d x$ and $\ell = x_{d} - x_{e}$. Using (\ref{evoop2}) as our neutrino evolution operator, the flavor eigenstates at a later time after the neutrino have traveled a distance $\ell$ are given by:
%+++++++++++++++++++++++++++++++++++++++++++++++++++++++++++++++++++++++++++++++
\begin{gather}
\left|\nu_{e}(\ell)\right>
=
i 
\sin \varphi(\ell)
\left(
\cos 2 \theta_{L}
\left|
\nu_{e}
\right>
-
e^{- i \sigma}\sin 2 \theta_{L}
\left|
\nu_{\mu}
\right>
\right)
+
\cos\varphi(\ell)
\left|
\nu_{e}
\right>
\\
\left|\nu_{\mu}(\ell)\right>
=
-
i
\sin \varphi(\ell)
\left(
e^{i \sigma} 
\sin 2 \theta_{L}
\left|
\nu_{e}
\right>
+
\cos 2 \theta_{L}
\left|
\nu_{\mu}
\right>
\right)
+
\cos\varphi(\ell)
\left|
\nu_{\mu}
\right>
\end{gather}
%+++++++++++++++++++++++++++++++++++++++++++++++++++++++++++++++++++++++++++++++
Therefore, if a neutrino of flavor $f$ is created and travels a distance $\ell$ to a detector, the probability that the neutrino is observed as a $f'$ flavored neutrino is 
%+++++++++++++++++++++++++++++++++++++++++++++++++++++++++++++++++++++++++++++++
\begin{gather}
P(\nu_{f} \rightarrow \nu_{f'})
=
|\left<
\nu_{f'}
\right|
\left.
\nu_{f}(\ell)
\right>|^{2}
=
\delta_{f,f'}
+
(-1)^{\delta_{f,f'}}
\sin^{2}2 \theta_{L}
\sin^{2} \varphi(\ell)
\end{gather}
%+++++++++++++++++++++++++++++++++++++++++++++++++++++++++++++++++++++++++++++++
It is interesting to consider the case when the k-essence field does not vary rapidly in space and time compared to terrestrial scales. In that case we can treat the integrand as a constant, and therefore the phase becomes  
%+++++++++++++++++++++++++++++++++++++++++++++++++++++++++++++++++++++++++++++++
\begin{gather}
\varphi(\ell)
=
\frac{|\Delta \tilde{m}^{2}|}{4 E}
\ell
\sqrt{1 + 2 y \cos 2 \Theta + y^{2}}
\end{gather}
%+++++++++++++++++++++++++++++++++++++++++++++++++++++++++++++++++++++++++++++++
Since the phase is proportional to the distance between the neutrino emitter and detector then we can write the oscillation probability as
%+++++++++++++++++++++++++++++++++++++++++++++++++++++++++++++++++++++++++++++++
\begin{gather}
P(\nu_{f} \rightarrow \nu_{f'})
=
\delta_{f,f'}
+
(-1)^{\delta_{f,f'}}
\sin^{2}2 \theta_{L}
\sin^{2}\frac{\pi l}{\lambda}
\label{OscProb}
\end{gather}
%+++++++++++++++++++++++++++++++++++++++++++++++++++++++++++++++++++++++++++++++
where $\lambda$ is the oscillation length and is defined as
%+++++++++++++++++++++++++++++++++++++++++++++++++++++++++++++++++++++++++++++++
\begin{gather}
\lambda
=
\frac{\pi \ell}{\varphi(\ell)}
=
\frac{4 \pi E}{|\Delta  \tilde{m}^{2}|}
\frac{1}{\sqrt{1 + 2 y \cos 2 \Theta + y^{2}}}
\end{gather}
%+++++++++++++++++++++++++++++++++++++++++++++++++++++++++++++++++++++++++++++++
It is interesting to compare results from the KINO mechanism with those from neutrino oscillations induced by mass. In the case where neutrino oscillations are due to mass entirely, the inverse oscillation length goes like  
%+++++++++++++++++++++++++++++++++++++++++++++++++++++++++++++++++++++++++++++++
\begin{gather}
4 \pi 
\lambda^{-1} = \frac{|\Delta  \tilde{m}^{2} |}{E}
\label{MassInducedLength}
\end{gather}
%+++++++++++++++++++++++++++++++++++++++++++++++++++++++++++++++++++++++++++++++
Likewise, if the mass of neutrinos vanishes then the inverse oscillation length goes like
%+++++++++++++++++++++++++++++++++++++++++++++++++++++++++++++++++++++++++++++++
\begin{gather}
4 \pi \lambda^{-1}
=
E
\left|
\Delta G
\right|
=
E
\left|
\frac{(g_{2} - g_{1}) u}{X}
\right|
(\dot{\phi}^{2} + \n \cdot \nabla \phi)^{2}
\label{KessenceInducedLength}
\end{gather}
%+++++++++++++++++++++++++++++++++++++++++++++++++++++++++++++++++++++++++++++++
Comparing (\ref{MassInducedLength}) and (\ref{KessenceInducedLength}) it is apparent that neutrino oscillations induced by either a flavor-non-diagonal mass term or a flavor-non-diagonal k-essence coupling will lead to noticeably different energy dependences for the oscillation length. If neutrino oscillations are entirely induced by k-essence then the oscillation length goes like $\lambda^{-1} \sim E$. The result (\ref{KessenceInducedLength}) should be compared to neutrino oscillations induced by either the VLI mechanism or VEP mechanism. Both the VLI and VEP mechanisms have the same $\lambda^{-1} \sim E$ behavior that the KINO mechanism has. This should come as no surprise since the flavor-dependent emergent metric $G_{\mu\nu}^{(\alpha)}$ can be viewed as a regular space-time metric but with a flavor dependent, and therefore equivalence principle violating, gravitational constant. Equations (\ref{OscProb}) and (\ref{KessenceInducedLength}) immediately tell us that in order for there to be flavor oscillations not only must $g_{1} \neq g_{2}$, but also $u \neq 0$. Since $u$ vanishes for $c_{s} =1$ this means that there must be lorentz violation if k-essence is to have any effect on neutrino oscillations. K-essence, therefore, acts as a lorentz violating ``aether'' background, and can be the motivation behind models of neutrino oscillation that invoke Lorentz violation.
%+++++++++++++++++++++++++++++++++++++++++++++++++++++++++++++++++++++++++++++++

%+++++++++++++++++++++++++++++++++++++++++++++++++++++++++++++++++++++++++++++++
Analysis of the available data from current and past neutrino observatories have tended to favor mass induced neutrino oscillations, which can produce the desired $\lambda \propto E$ type behavior. Therefore, the KINO mechanism alone does not suffice to explain the observations seen in the numerous neutrino experiments that have been carried out. However, it is still possible that k-essence could be a subleading contribution to neutrino oscillations with mass being the dominant cause. Studies have looked into the possibility of alterations to the leading order $\lambda \propto E$ dependence and have been able to place very tight constraints on the coefficients of sub-leading contribution to the energy dependence of $\lambda$. In \cite{Fogli:1999fs} they considered the possibility of different mechanisms inducing oscillations in the $\nu_{\mu} \leftrightarrow \nu_{\tau}$ channel, among which were the VLI and VEP mechanisms. Since the VLI and VEP scenarios both lead to the same energy dependence for the oscillation length that KINO does, their constraints on the VLI and VEP coefficients can be easily translated to a bound on $\Delta G$
%+++++++++++++++++++++++++++++++++++++++++++++++++++++++++++++++++++++++++++++++
\begin{gather}
|\Delta G|
<
1.2
\times
10^{-23}
\end{gather}
%+++++++++++++++++++++++++++++++++++++++++++++++++++++++++++++++++++++++++++++++
this is the most conservative bound that can be placed on $|\Delta G|$ and is independent of the mixing angle $\theta_{K}$. This bound would seem to cast doubt on KINO as an even subleading effect in the $\nu_{\mu} \leftrightarrow \nu_{\tau}$ channel. We would like to emphasize that this is a strong indication of the nature of the coupling of neutrinos to the k-essence scalar. Our analysis suggests that if the k-essence scalar field exists, in order to be phenomenologically viable, its couplings must be flavor blind. 
%+++++++++++++++++++++++++++++++++++++++++++++++++++++++++++++++++++++++++++++++

%+++++++++++++++++++++++++++++++++++++++++++++++++++++++++++++++++++++++++++++++
Although KINO would seem to be immediately discounted from consideration, it may be possible to realistically consider this model if we are ready to include further symmetry violating terms in our action. It has been shown in \cite{Katori:2006mz} that if lorentz {\it and certain types of CPT} violating terms are included in the neutrino action then it becomes possible to create {\it pseudo-mass} terms at high energies just when lorentz violation effects should be taking over. These types of models open up the possibility of a unified explanation for all the existing neutrino data, including the controversial LSND results \cite{Athanassopoulos:1997pv}. Since these models have been at least qualitatively compatible with experiment, this leaves open the possibility that k-essence could still play a role in neutrino oscillations.

As we have just mentioned this approach requires a specific kind of CPT violating term in the neutrino action.   Although it is always possible to put terms into the action arbitrarily, in principle it might be possible for k-essence to be a source for these as well. K-essence could lead to the needed CPT breaking terms in the neutrino action by one of two ways, the first is by a possible axial vector term in the spinor covariant derivative\footnote{See appendix \ref{SpinCon}}. In this paper we have assumed that space-time is flat, and as a result the covariant derivative is proportional to $\gamma^{\mu}$. However, if more general curved space-times are considered then in general the spinor connection will have a non-zero axial vector part. Another way in which k-essence can source the necessary CPT violating term is by considering the possibility of torsion in the emergent space-time. Such a treatment of k-essence would require taking into account of the spin-orbit coupling between the neutrino and the k-essence field and treating the metric and connection on the emergent space-time as independent variables each requiring their own field equations. Torsion in k-essence backgrounds is an interesting possibility and deserves further study in its own right.

%+++++++++++++++++++++++++++++++++++++++++++++++++++++++++++++++++++++++++++++++

%+++++++++++++++++++++++++++++++++++++++++++++++++++++++++++++++++++++++++++++++
%+++++++++++++++++++++++++++++++++++++++++++++++++++++++++++++++++++++++++++++++
%+++++++++++++++++++++++++++++++++++++++++++++++++++++++++++++++++++++++++++++++
%+++++++++++++++++++++++++++++++++++++++++++++++++++++++++++++++++++++++++++++++
\section{Conclusion}
\label{Conclusions}
%+++++++++++++++++++++++++++++++++++++++++++++++++++++++++++++++++++++++++++++++
%+++++++++++++++++++++++++++++++++++++++++++++++++++++++++++++++++++++++++++++++
%+++++++++++++++++++++++++++++++++++++++++++++++++++++++++++++++++++++++++++++++
%+++++++++++++++++++++++++++++++++++++++++++++++++++++++++++++++++++++++++++++++
In this paper we have investigated the effects on neutrino velocity and oscillations, when neutrinos couple to a k-essence field background though the emergent metric $G_{\mu\nu}$. Specifically we have studied the results of replacing the vierbein $e_{a}^{\mu}$ of the gravitational background with the vierbein $E_{a}^{\mu}$ of the emergent k-essence geometry in the neutrino action. The first implication of this coupling is a change in the neutrino dispersion relation, which means that the neutrino velocity would be dependent on the k-essence background in which it propagates. Without k-essence massless neutrinos always propagate at the speed of light with respect to local observers. However, with k-essence, massless neutrinos will move at a new speed $c_{\nu}$ with respect to an observer who perceives a uniform k-essence background. Therefore, if future observations show that neutrinos do travel at less than the speed of light, it cannot be conclusively determined by this alone if this is due to neutrinos being massive or if neutrinos are massless but coupled to k-essence. However, at present, no observations have found a measurable difference between the speed of light and neutrinos. If neutrinos are assumed to be massless, than past observations constrain the deviation of $c_{\nu}$ from the speed of light to no more than 1 part in $10^{8}$.
%+++++++++++++++++++++++++++++++++++++++++++++++++++++++++++++++++++++++++++++++

%+++++++++++++++++++++++++++++++++++++++++++++++++++++++++++++++++++++++++++++++
The other effect that k-essence has on neutrinos is in the phenomenon of neutrino oscillations. We have shown  that if the k-essence-neutrino coupling $g$ is promoted to the status of an operator $\hat{g}$ that acts in the neutrino flavor space, neutrino oscillations are produced even in the absence of a neutrino mass term. We found that if neutrino oscillations are caused entirely by k-essence, then the oscillation length depends on the energy like $\lambda \sim E^{-1}$. This is to be contrasted with the result from mass induced neutrino oscillations where $\lambda \sim E$. Thus, while neutrino oscillations can be induced by k-essence, it will lead to a drastically different observation of the energy dependence of the oscillation length. Current data seems to favor a $\lambda \sim E$ behavior which implies an important constraint on the couplings of neutrinos to the k-essence scalar. In order to be phenomenologically viable these couplings must be flavor diagonal. Our analysis is a very good example of how interactions between dark energy and visible matter can be used to constrain their couplings.

%+++++++++++++++++++++++++++++++++++++++++++++++++++++++++++++++++++++++++++++++

%+++++++++++++++++++++++++++++++++++++++++++++++++++++++++++++++++++++++++++++++
Neutrino oscillations induced by k-essence have many of the same properties that some \cite{Gasperini:1988zf,Halprin:1991gs,Coleman:1997xq}
 earlier proposed mechanisms had, and in fact k-essence can be seen as a realization of these past phenomenologically motivated models. In VLI models the lorentz violation is incorporated by introducing some preferred 4-vector into the neutrino action. This preferred 4-vector has the interpretation in k-essence as the 4-gradient of the k-essence field. In fact, as can be seen from the formulae in section \S\ref{NeutrinoOscillations}, the physical effects vanish at $c_s =1$ where our model has exact Lorentz invariance. The VEP mechanism attempts to explain the source of neutrino oscillations as a consequence of a flavor-non-diagonal coupling to gravity. Although perhaps correct from a theoretical point of view, this mechanism calls for a very drastic change in our understanding of fundamental physics; namely it requires us to give up the long held notion of the equivalence principle. The type of coupling studied in this paper mimics exactly the VEP mechanism, but since the flavor non-diagonal coupling is in the k-essence sector instead of the the gravitational sector, the equivalence principle is maintained. In short, k-essence can be seen as the source of lorentz violation in the VLI model, and alternatively, a reinterpretation of the VEP model. KINO can therefore be seen as a theoretically and phenomenologically motivated manifestation of earlier attempts at alternatives to mass-induced neutrinos oscillations.
 
 In this paper, we have coupled the neutrinos to the k-essence scalar using the emergent metric which is covariantly constant. In a future work we will show how this can be generalized to more general couplings by introducing the analogue of torsion for the emergent space-time. This generalized coupling will allow us to analyze the role, if any, of CPT violation in the coupling of k-essence to fermions. If CPT violation can be naturally produced, then it would be possible to discuss more realistic models of neutrino oscillations as discussed in \cite{Katori:2006mz}.
 
 %+++++++++++++++++++++++++++++++++++++++++++++++++++++++++++++++++++++++++++++++

%+++++++++++++++++++++++++++++++++++++++++++++++++++++++++++++++++++++++++++++++
\section{Acknowledgments}
%+++++++++++++++++++++++++++++++++++++++++++++++++++++++++++++++++++++++++++++++
The authors would like to thank E. Babichev, A. Dolgov, B. Roe and A. Vikman for their helpful discussion and comments. R.A. would like to thank P. Binetruy and the theory group at APC Paris-7, for their hospitality and for discussions while this work was in progress.
 %+++++++++++++++++++++++++++++++++++++++++++++++++++++++++++++++++++++++++++++++

%+++++++++++++++++++++++++++++++++++++++++++++++++++++++++++++++++++++++++++++++
%+++++++++++++++++++++++++++++++++++++++++++++++++++++++++++++++++++++++++++++++
%+++++++++++++++++++++++++++++++++++++++++++++++++++++++++++++++++++++++++++++++
%+++++++++++++++++++++++++++++++++++++++++++++++++++++++++++++++++++++++++++++++
\appendix
\section{Dirac Fermions in the Emergent K-essence Metric}
\label{SpinCon}
%+++++++++++++++++++++++++++++++++++++++++++++++++++++++++++++++++++++++++++++++
%+++++++++++++++++++++++++++++++++++++++++++++++++++++++++++++++++++++++++++++++
%+++++++++++++++++++++++++++++++++++++++++++++++++++++++++++++++++++++++++++++++
%+++++++++++++++++++++++++++++++++++++++++++++++++++++++++++++++++++++++++++++++
%+++++++++++++++++++++++++++++++++++++++++++++++++++++++++++++++++++++++++++++++
%+++++++++++++++++++++++++++++++++++++++++++++++++++++++++++++++++++++++++++++++
The simplest action of a Dirac fermion $\psi$ coupled to a metric $G_{\mu\nu}$ is conventionally written as
%+++++++++++++++++++++++++++++++++++++++++++++++++++++++++++++++++++++++++++++++
\begin{gather}
S
=
\int
d^{4}x
E
\bar{\psi}
\left[
i \gamma^{a}
E_{a}^{\mu}
\D_{\mu}
-
M
\right]
\psi
\end{gather}
%+++++++++++++++++++++++++++++++++++++++++++++++++++++++++++++++++++++++++++++++
where $E_{a}^{\mu}$ is the vierbein corresponding to the metric $G_{\mu\nu}$, and $\D_{\mu}$ is the spinor covariant derivative which is given by
%+++++++++++++++++++++++++++++++++++++++++++++++++++++++++++++++++++++++++++++++
\begin{gather}
\D_{\mu}
=
\partial_{\mu}
-
\frac{i}{4}
\Omega_{ab\mu}
\sigma^{ab}.
\end{gather}
%+++++++++++++++++++++++++++++++++++++++++++++++++++++++++++++++++++++++++++++++
Here $\sigma^{ab} = \frac{i}{2}\left[\gamma^{a} , \gamma^{b}\right]$ and $\Omega_{ab\mu}$ is the spinor connection form which by definition is
%+++++++++++++++++++++++++++++++++++++++++++++++++++++++++++++++++++++++++++++++
\begin{gather}
\Omega_{ab\mu}
=
E_{a\nu}
\partial_{\mu}
E^{\nu}_{b}
+
E_{a\nu}
E^{\sigma}_{b}
\Gamma^{\nu}_{\sigma \mu}
\end{gather}
%+++++++++++++++++++++++++++++++++++++++++++++++++++++++++++++++++++++++++++++++
where $\Gamma^{\nu}_{\sigma \mu}$ is the standard Christoffel symbol. After some work, the spinor covariant derivative can be shown to satisfy
%+++++++++++++++++++++++++++++++++++++++++++++++++++++++++++++++++++++++++++++++
\begin{gather}
\slash{\D}
=
\gamma^{a}
E^{\mu}_{a}
\D_{\mu}
=
\gamma^{a}
E^{\mu}_{a}
\left[
\partial_{\mu}
+
\frac{1}{2}
E^{\nu}_{b}
\left(
\partial_{\mu}
E_{\nu}^{b}
-
\partial_{\nu}
E_{\mu}^{b}
\right)
+
\frac{i}{2}
\gamma_{5}
A_{ \mu}
\right]
\label{CovariantDer}
\end{gather}
%+++++++++++++++++++++++++++++++++++++++++++++++++++++++++++++++++++++++++++++++
where
%+++++++++++++++++++++++++++++++++++++++++++++++++++++++++++++++++++++++++++++++
\begin{gather}
A_{\mu}
=
\frac{1}{4}
\epsilon^{a b c d}
E_{a\mu}
\left(
\partial_{\sigma}
E_{b \nu}
-
\partial_{\nu}
E_{b \sigma}
\right)
E_{c}^{\nu}
E_{d}^{\sigma}
\end{gather}
%+++++++++++++++++++++++++++++++++++++++++++++++++++++++++++++++++++++++++++++++
The formula (\ref{CovariantDer}) is valid not just for the emergent geometry of k-essence, but for all geometries with a vierbein. In this paper we will assume that the space-time geometry is flat. In this case the axial vector part vanishes since $A_{\mu} =0$ always if space-time is flat, and the only non-zero part of the spin connection is the vector portion. Therefore, the spinor covariant derivative for a general k-essence field and a flat space-time background is 
%+++++++++++++++++++++++++++++++++++++++++++++++++++++++++++++++++++++++++++++++
\begin{gather}
\slash{\D}
=
\Omega
\gamma^{a}
\delta_{a}^{\mu}
\Bigg[
\partial_{\mu}
+
\frac{g u}{2 X}
\partial_{\mu}
\phi
(\partial^{\nu}\phi
\partial_{\nu})
+
\frac{g u}{4 X}
\partial_{\mu}
\phi
\partial^{2}
\phi
-
\frac{g \partial_{\nu} u}{2 (1 + g u)}
\left(
\delta_{\mu}^{\nu}
-
\frac{\partial^{\nu}\phi \partial_{\mu}\phi}{2 X}
\right)
\nn \\
+
\frac{g u}{4 X}
\partial_{\nu}X
\left(
\delta_{\mu}^{\nu}
-
\frac{\partial^{\nu}\phi \partial_{\mu}\phi}{X}
\right)
-
\frac{3}{2} \frac{\partial_{\nu}\Omega}{\Omega}
\left(
\delta^{\nu}_{\mu}
+
\frac{gu}{2 X}
\partial^{\nu}
\phi
\partial_{\mu}
\phi
\right)
\Bigg]
\end{gather}
%+++++++++++++++++++++++++++++++++++++++++++++++++++++++++++++++++++++++++++++++
It the cosmologically relevant case where $\phi$ is time dependent but $\nabla_{i}\phi =0$, the covariant derivative in this case leads to 
%+++++++++++++++++++++++++++++++++++++++++++++++++++++++++++++++++++++++++++++++
\begin{gather}
\slash{\D}
=
\Omega
\gamma^{a}
\delta_{a}^{\mu}
\left[
\partial_{\mu}
+
\delta_{\mu}^{ 0}
\left(
g u
\partial_{0}
-
(1 + g u)
\frac{3 \dot{\Omega}}{2 \Omega}
\right)
\right]
\end{gather}
%+++++++++++++++++++++++++++++++++++++++++++++++++++++++++++++++++++++++++++++++
In k-essence models that attempt to explain the cosmological constant problem, the higher order derivatives of the field become irrelevant at late times since the field reaches a steady state by that point. Therefore, it is reasonable to ignore the spinor connection in our analysis. However, if one is interested in the effect k-essence has on neutrinos in the early universe, in particular around the time of matter-radiation equality, then the spinor connection can be an important contribution to the neutrino action.  
%+++++++++++++++++++++++++++++++++++++++++++++++++++++++++++++++++++++++++++++++

\bibliographystyle{h-elsevier}
\bibliography{Neutrino-K-Essence}

\end{document}